\documentclass[prb,twocolumn,english,superscriptaddress]{revtex4-1}
\usepackage{amsmath,amssymb,mathrsfs}

\usepackage{natbib}
\usepackage{subfigure}
\usepackage{tabularx}
\usepackage{epsfig}
\usepackage{longtable}
\usepackage{amsfonts}
\usepackage{rotating}
\usepackage{subfigure}
\usepackage{amsmath}
\usepackage{babel}
\usepackage{comment}

\usepackage[unicode=true,
 bookmarks=true,bookmarksnumbered=false,bookmarksopen=false,
 breaklinks=false,pdfborder={0 0 1},backref=false,colorlinks=true]
 {hyperref}
\hypersetup{
linkcolor=magenta, urlcolor=blue, citecolor=blue, pdfstartview={FitH}, hyperfootnotes=false, unicode=true}

\usepackage{color}

\def\nn{\nonumber}

\newcommand{\bea}{\begin{eqnarray}}
\newcommand{\eea}{\end{eqnarray}}

\newcommand{\be}{\begin{equation}}
\newcommand{\ee}{\end{equation}}



\begin{document}
\title{ Quantum nonergodicity  and fermion localization in a system with a single-particle mobility edge} 
\author{Xiaopeng Li}
\email{xiaopeng@umd.edu} 
\affiliation{Condensed Matter Theory Center and Joint Quantum Institute, Department
of Physics, University of Maryland, College Park, MD 20742-4111, USA}
\author{J. H. Pixley}
\affiliation{Condensed Matter Theory Center and Joint Quantum Institute, Department
of Physics, University of Maryland, College Park, MD 20742-4111, USA}
\author{Dong-Ling Deng} 
\affiliation{Condensed Matter Theory Center and Joint Quantum Institute, Department
of Physics, University of Maryland, College Park, MD 20742-4111, USA}
\author{Sriram Ganeshan}
\affiliation{Condensed Matter Theory Center and Joint Quantum Institute, Department
of Physics, University of Maryland, College Park, MD 20742-4111, USA}
\affiliation{Simons Center for Geometry and Physics, Stony Brook, NY 11794, USA}

\author{S. Das Sarma} 
\affiliation{Condensed Matter Theory Center and Joint Quantum Institute, Department
of Physics, University of Maryland, College Park, MD 20742-4111, USA}

\begin{abstract}
We study the many-body localization aspects of single-particle mobility edges in fermionic systems. We investigate incommensurate lattices and random disorder Anderson models. Many-body localization and quantum nonergodic properties are studied by comparing entanglement and thermal entropy, and by calculating the scaling of subsystem particle number fluctuations, respectively. We establish a nonergodic extended phase as a  generic 
intermediate phase (between purely ergodic extended and nonergodic localized phases)
for the many-body localization transition of non-interacting fermions
where the entanglement entropy manifests a volume law (hence, `extended'), but there are large fluctuations in the subsystem particle numbers (hence, `nonergodic'). 
{Based on the numerical results, we expect} 
such an intermediate phase
scenario 
may 
continue to hold even for the many-body localization in the presence of interactions
as well. We find for many-body fermionic states in non-interacting one dimensional Aubry-Andr\'e and three dimensional Anderson models that the entanglement entropy density and the normalized particle-number fluctuation  have discontinuous jumps at the localization transition 
where the entanglement entropy is sub-thermal but obeys the ``volume law". In the vicinity of the localization transition we find that both the entanglement entropy and the particle number fluctuations 
obey a
single parameter scaling based on the diverging localization length. 
We argue using numerical and theoretical results that such a critical scaling behavior should persist for the interacting many-body localization problem with important observable consequences.  Our work provides persuasive evidence in favor of there being two  
{transitions} 
in many-body systems with single-particle mobility edges, the first one indicating a transition from the purely localized nonergodic many-body localized phase to a nonergodic extended many-body metallic phase, and the second one being a transition eventually to the usual ergodic many-body extended phase.
\end{abstract}
\maketitle

\section{Introduction} 

Quantum thermalization of isolated  systems undergoing unitary time evolution is a far-reaching fundamental problem in quantum statistical mechanics. 
The issue of whether a large quantum system can act as its own bath touches upon the deepest questions in statistical mechanics such as the equivalence between microcanonical (``constant energy") and canonical (``constant temperature") ensembles, and even the applicability of the concepts of thermodynamic equilibrium and temperature in describing isolated  quantum systems. 
To reconcile unitarity and quantum ergodicity, it has been conjectured that typical isolated quantum systems  will thermalize in the long-time limit,  in a sense of the so-called eigenstate thermalization hypothesis (ETH)~\cite{1991_Deutsch_ETH_PRA,1994_Srednicki_ETH_PRE}, where the neighborhood of a subsystem acts as an effective quantum bath, 
making microcanonical and canonical ensembles equivalent for large isolated systems.  In fact, the ETH hypothesis in its strong form asserts that every eigenstate of an isolated quantum system is thermal.  Whether this is true generically for all interacting quantum systems in the presence of disorder and interaction is a topic of great current theoretical interest. 
Understanding quantum ergodicity becomes more relevant 
(and no longer just a purely abstract theoretical question)
with the recent experimental progress  in synthesizing artificial many-body systems, e.g., with atoms, photons, or artificial qubits, where couplings to the environment could be made negligible~\cite{2008_Bloch_RMP,2013_Esslinger_RMP,2010_Singer_RMP,2013_Xiang_RMP}, 
and therefore, the issue of an ``isolated" quantum system is no longer just an abstraction and can be tested in the laboratory.  Current activity in building a fault-tolerant quantum computer, which necessitates a great degree of isolation for the system from the environment, also raises the question of the fundamental statistical mechanics of isolated interacting quantum systems. 
Although the conjecture of quantum ergodicity has not been proved for generic systems, it  has been confirmed in a few cases by numerical studies for clean many-body systems~\cite{2008_Rigol_thermalization_Nature,2008_Rigol_PRA}. For certain integrable models including free fermions, it has been shown that the local observables of typical eigenstates respect ETH~\cite{2011_Gogolin_ETH_PRL,2013_Ueda_ETH_PRE,2015_Yang_ETH,2015_Alba_ETH_PRB,2015_Magan_FreeFermion_arXiv,2016_Gluza_Gaussification_arXiv}. 
This is a non-trivial statement as the conserved quantities in integrable models do not satisfy ETH 
(and one would have thought that a free system can never thermalize!) in any sense. 
For such thermal systems, the entanglement entropy is an extensive quantity 
(often referred to as ``the volume law" since the entanglement, being extensive, grows as the volume of the system in contrast to ground states of quantum systems where the entanglement usually obeys ``an area law" up to logarithmic corrections-- this distinction between volume and area laws for the entanglement entropy turns out to be a key quantity of interest in studying quantum thermalization of quantum systems). 

As a function of disorder strength, isolated quantum systems can undergo a transition from an extended thermal phase to  a many-body localized (MBL) nonthermal phase~\cite{AndersonLocalization,Basko06,huse2015many,2015_Altman_review}, where a subsystem would fail to thermalize with its  neighborhood~\cite{2011_Canovi_Rossini_MBL_PRB,2014_Huse_MBL_PRB,huse2015many,2015_Altman_review}. The existence of the MBL phase in interacting disordered systems has been demonstrated by perturbative arguments~\cite{Basko06}, which was  then further confirmed by recent numerical work~\cite{oganesyan2007,2008_Marko_XXZMBL_PRB,pal2010,moore2012,vadim,bardarson2014,2015_Li_MBL_PRL,modak2015many,2015_Luitz_MBLedge_PRB,2015_Mondragon_arXiv,2015_Moure_MBL} and a mathematical proof~\cite{imbrie2014many}.  To locate the MBL transition, different aspects have been examined~\cite{2007_Burrell_PRL,moore2012,2011_Canovi_Rossini_MBL_PRB,2013_Serbyn_PRL,2014_Nanduri_Kim_PRB,2014_Pekker_MBL_PRX,2014_Kwasigroch_Cooper_MBL_PRA,2014_Serbyn_Knap_MBL_PRL,2014_Rahul_MBL_PRB,2014_Yao_Laumann_PRL,2014_Chandran_MBL_PRB,2014_Grover_MBL_arXiv,2014_Grover_Fisher_QL,2014_Andraschko_MBL_PRL,2014_Laumann_Pal_PRL,2014_Hickey_MBLglass_arXiv,2014_Lev_MBL_PRB,2015_Agarwal_MBL_PRL,2015_Vosk_MBL_PRX,2015_Potter_MBL_PRX,2015_Singh_MBL_PRL,2015_Mueller_arXiv,2015_Deng_OC_arXiv,2015_Gopalakrishnan_PRB,2015_Johri_MBL_PRL,2015_Ponte_Abanin_PRL,2015_You_Qi_arXiv,2015_Singh_Bardarson_arXiv,2015_Lev_arXiv,2015_Chen_arXiv,2015_Baygan_Lim_arXiv,2015_Yu_Pekker_arXiv,2015_Khemani_Pollmann_arXiv,2015_Galitski_MBL_arXiv,2015_Wang_Chen_arXiv,2015_Chandran_Laumann_arXiv,2015_Burin_MBL_PRB}, for example, entanglement entropy, level repulsion, and non-thermal fluctuations.  Across the transition, the entanglement entropy becomes an intensive quantity~\cite{2013_Bauer_JSM}  
(i.e. changing from a volume to an area law), 
thus no longer respecting the thermal entropy;  the level repulsion disappears 
in the MBL phase
giving  rise to Poisson level statistics; and the non-thermal fluctuations are greatly enhanced violating ETH~\cite{2015_Li_MBL_PRL} 
since the system is no longer in thermodynamic equilibrium by itself. It is however unclear how to relate the transitions determined from these different MBL diagnostics. 
Are all of them always exactly equivalent (as has almost always been assumed, but never established explicitly) with respect to the MBL transition?  (It is already known that one particular diagnostic, namely, the inverse participation ratio, which is used extensively to study single-particle localization is a poor indicator for the MBL transition although it is very efficient in identifying the single-particle Anderson localization.) 
Recent theoretical arguments and numerical studies suggest that the localization and nonthermal transitions may not always coincide~\cite{2014_Lev_MBL_PRB,2015_Li_MBL_PRL,2015_Lim_Sheng_arXiv,2015_Pino_Altshuler_NonErgodicMetal}. 
Studying this question in depth for fermionic systems with single-particle mobility edges (separating the energy eigenstates sharply between localized and extended) is the goal of the current work.

When considering many-body mobility edge physics (i.e. investigating MBL in a system whose noninteracting counterpart has a single-particle mobility edge), understanding the possible scenarios of MBL transitions faces additional complications~\cite{2014_Laumann_Pal_PRL,2015_Li_MBL_PRL,2015_Huang_MBL_arXiv,2015_Baygan_Lim_arXiv,2015_Muller_arXiv,modak2015many}. In the context of single-particle localization, the mobility edge is well-established in the three dimensional Anderson model~\cite{AndersonLocalization} and also various one dimensional incommensurate lattice models~\cite{AA, azbel, harper55,griniasty1988localization, sankarprl88, thouless1988localization, sankarprb90, 2010_Biddle_PRL, biddleprb11, sriramgaa}. In the MBL context, resolving the mobility edge in a fully interacting system is very challenging, 
particularly using purely numerical techniques because of severe finite size effects. 
Nonetheless, some progress has been made in recent numerical studies and some evidence 
for the existence of the
many-body mobility edge has been presented. But a theoretical consensus about the thermodynamic limit is still lacking due to the limitation of simulating large interacting systems~\cite{2015_Muller_arXiv}. 
In fact, an extreme view has been that for systems with single-particle mobility edges, the corresponding interacting systems will exhibit no MBL and will be manifestly thermal.  The argument for this extreme view, which, in our opinion, is ill-founded, is that such a system must always have mixed extended and localized single-particle orbitals forming the many-body wavefunctions, leading to thermalization explicitly obeying ETH.  Aside from the fact that, if this extreme view were valid, then MBL becomes a subject only of very limited academic interest since generic three dimensional disordered systems typically have single-particle mobility edges, we believe that the mixing of the localized and extended single-particle orbitals (at different single-particle energies) in forming the many-body states could in fact lead to the possibility of the existence of an intermediate phase which is `extended' (by virtue of the extended orbitals contributing to the entanglement entropy) but nonthermal  (by virtue of the localized orbitals contributing to large nonthermal fluctuations).   Recent numerical work provides evidence for such an intermediate MBL phase~\cite{2015_Li_MBL_PRL,modak2015many,2015_Rahul_PRB,2016_Bauer_arXiv}, 
and in our current work, we study this problem in great depth providing possible scenarios leading to this exotic nonthermal metallic phase, intermediate between a pure thermal and a pure MBL phase.

In this work, we investigate thermalization and localization properties of non-interacting fermions through a one dimensional (1d) incommensurate lattice model and the three dimensional (3d) Anderson model.  
(Both classes of models we study have single-particle mobility edges sharply separating the one particle energy eigenstates into localized and extended states.)
We calculate the entanglement entropy  scaling and the subsystem particle number fluctuations to track the localization and nonergodic transitions, respectively. 
In the following manuscript, we study the thermal to non-thermal transition in the context of local observables of a subsystem, which can satisfy ETH (despite the model being non-interacting) as described in detail in Secs.~\ref{sec:quadraticfluc} and \ref{sec:Diagnostics}.
In a completely  extended phase, we find  that the entanglement entropy obeys the volume-law and that non-thermal fluctuations are suppressed, and consequently the system is in a thermal phase whose thermal entropy is equal to the entanglement entropy. In a completely localized phase, we find that the entanglement entropy follows an area law and non-thermal fluctuations exhibit a  $\sqrt{N}$ scaling (for particle number $N$).  
(These two phases, thermal and localized, are of course already known in the literature through studies of models with no single-particle mobility edges.)
We establish the scaling relation between the particle number fluctuation and the localization length in a many-body localized phase. 
 In the presence of a single-particle mobility edge, we find that the interplay of localized and delocalized degrees of freedom leads to an intermediate phase where we have both volume-law entanglement entropy (with a sub-thermal value) and strong non-thermal fluctuations~\cite{2015_Li_MBL_PRL}. In this phase, entanglement and thermal entropies, both being extensive, strongly deviate from each other. At the localization transition, we find for 1d Aubry-Andr\'e (AA) and 3d Anderson models that the entanglement entropy has discontinuous jumps in the thermodynamic limit.  For all the investigated models with a mobility edge, we find a considerable parameter region where the entanglement entropy is extensive but  does not match the thermal entropy. 
Our results for particle number fluctuations give important guidelines on how to experimentally distinguish nonergodicity and localization, which we do not find to be equivalent in systems with single-particle mobility edges.  Our discovery of a distinction between localization and nonergodicity (i.e. they do not necessarily coincide) is a result of fundamental significance, which has not been appreciated before (except in the context of Refs.~\onlinecite{2015_Li_MBL_PRL,modak2015many,2015_Rahul_PRB,2016_Bauer_arXiv}).

The goal of the current work is to study MBL properties of non-interacting (although we do present some results for the interacting systems too, see, e.g., Figs.~\ref{fig:1DIntAA} and~\ref{fig:MBDOS}) many-body systems, where the corresponding single-particle problem has a mobility edge, in order to clearly understand (and establish) the emergence of an intermediate (i.e. in between the ergodic extended phase and the non-ergodic many-body-localized phase) non-ergodic `metallic' phase.  We emphasize that recent work in interacting systems has indicated that MBL exists in interacting systems which have single-particle mobility edges, and more interestingly, there is an intermediate phase which is simultaneously nonergodic and extended in such a system in the presence of interaction~\cite{2015_Li_MBL_PRL}. Since the studied interacting MBL systems in Ref.~\onlinecite{2015_Li_MBL_PRL}  have rather limited system sizes ($10$-$20$ particles), it is important to investigate finite size effects by going to much larger systems in order to ensure the robustness of the MBL and the intermediate phase, which is obviously impossible in an interacting system where the many-body Hilbert space becomes prohibitively large for $>20$ interacting particles.  This motivates the current study, where using noninteracting many-body states we can numerically study large systems ($\sim 100$ or more particles), to verify the earlier interacting small system study results~\cite{2015_Li_MBL_PRL,modak2015many}.  An important aspect of the current study is that the intermediate non-ergodic extended phase (in between purely non-ergodic localized and ergodic extended phases) emerges naturally here as `mixed' many-body states formed by combining both extended and localized single-particle orbitals of the corresponding one-particle Schrodinger equation.  The current work thus establishes the existence of the intermediate non-ergodic extended phase definitively without any finite-size constraint.  Since the existence of such an intermediate phase has already been reported~\cite{2015_Li_MBL_PRL} in the interacting many-body problem (albeit in small systems), it is reasonable to expect that the intermediate non-ergodic extended phase is a generic phenomenon in the thermodynamic limit which remains stable in the presence of finite interactions.

In order to avoid any misunderstanding or confusion, we mention right in this introduction what our paper is not about.  It is not about {\it many-body-mobility-edges}, which is a distinct topic of considerable interest and importance by itself.  We do not discuss or even consider the question of whether many body localization itself could lead to the emergence of many body mobility edges.  All currently existing many body localization studies (except for Refs.~\onlinecite{2015_Li_MBL_PRL,modak2015many}, and the current work) start with noninteracting systems where all single-particle states are localized (either by random disorder as in the Anderson mode or by deterministic aperiodicity in the lattice potential as in the Aubry-Andr\'e model), and then explore what happens to the many-body eigenstate spectra as the interaction is turned on.  If the many body spectrum remains localized (as reflected in the entanglement entropy being an area law) up to a finite interaction strength (for a fixed disorder), then the system is considered to manifest many body localization, and the the many body localization transition is characterized by the critical interaction strength (in units of disorder strength) where the many body spectrum changes from being localized (area law) to delocalized (volume law).  It is, of course, in principle possible that the many body spectrum itself exhibits a many body mobility edge as a function of the interaction strength (i.e. the spectrum is localized up to a specific many body energy density and delocalized above that) even though the corresponding noninteracting single-particle spectrum is strictly localized everywhere.  This would be the analog of the single-particle mobility edge in the many-body localization problem where the effective mobility edge, created entirely by interaction effects, divides the many body eigenstates into localized and delocalized parts according to their energy densities.  Our work does not have anything to say about this important issue of the existence or not of the interaction-driven many body mobility edge.  We start from a noninteracting fermionic problem where the single-particle spectrum has an explicit mobility edge, and then ask whether the corresponding many-body spectrum could manifest localization, and if so, what the nature of this localization transition is, finding that the many-body spectrum in the presence of single-particle mobility edges allow for a new kind of `intermediate' nonergodic delocalized phase where the entanglement entropy obeys a subthermal volume law while at the same time thermodynamic quantities manifest large nonthermal fluctuations.

This paper is organized as follows: In section~\ref{sec:quadraticfluc} we introduce the notion of fluctuations of one-body observables across nearby eigenstates. In section~\ref{sec:Diagnostics} we establish the diagnostics to study the localization and the thermalization transition separately. In sections \ref{sec:AAtransition} and \ref{sec:MEtransition} we study the localization/thermalization transitions in the absence and presence of a mobility edge respectively. 
In section~\ref{sec:singleparticle}, 
which is a self-contained section independent of the rest of the paper providing the critical properties and the local integrals of motion for the incommensurate localization models we study,
we discuss level statistics, critical properties, and local integrals of motion of the incommensurate lattice model with a single-particle mobility edge. We conclude the paper with a summary of our main results and a discussion of the open questions in section~\ref{sec:summary}.

\section{Quadratic Hamiltonian and strong fluctuations of one-body observables}
\label{sec:quadraticfluc} 

\begin{figure}[htp!] 
\includegraphics[angle=0,width=.8\linewidth]{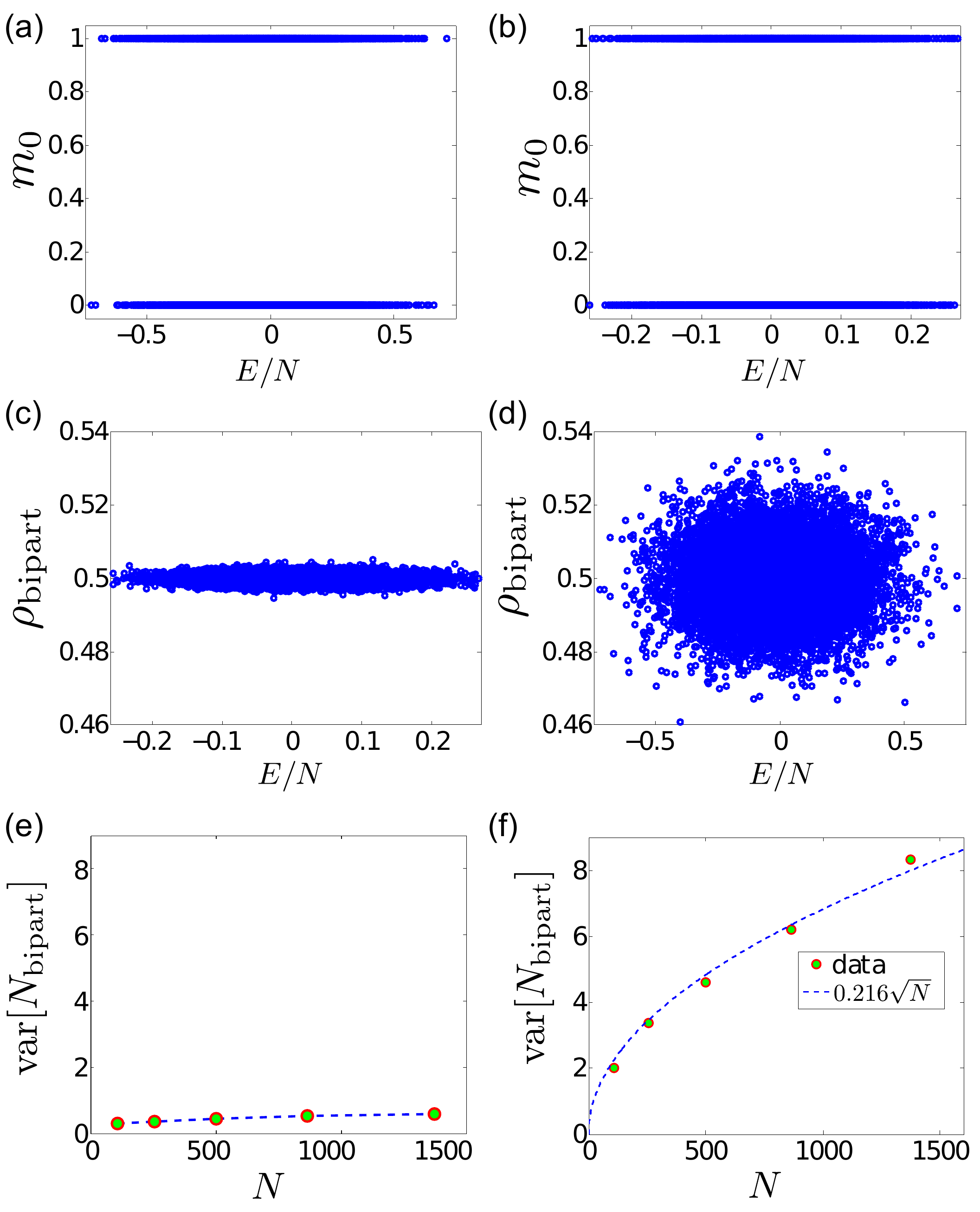}
 \caption{Many-body states in the 3d Anderson model. Here we randomly sample $10^5$ many-body eigenstates  of fermions at half filling. In (a), (c), and (e), we choose a disorder strength $W/t = 3$, and sample extended states. In (b), (d), and (f), we choose $W/t = 20$ with all states localized. (a) and (b) show the occupation number $m_0$ of the eigenmode at the single-particle spectra center. The occupation numbers for other orbitals behave similarly. For both extended and localized systems, we have strong fluctuations in $m_0$. (c) and (d) show the bipartite 
 ($\gamma = \frac{1}{2}$) density $\rho_{\rm bipart} = \frac{ N_{\rm bipart} } {D/2}$, with $N_{\rm bipart}$ the particle number in one half of the system. For localized states, the bipartite density exhibits strong fluctuation as shown in (d), whereas it is suppressed for extended states as shown in (c). (e) and (f) show the scaling of the fluctuation with increasing system size. In (e) and (f) we average over energy, since the energy dependence of the fluctuation is negligible. The numerical results agree with our analytical analysis (Eq.~\eqref{eq:Nsubscaling}). }
\label{fig:3DAnderson} 
\end{figure}

\begin{figure}[htp!] 
\includegraphics[angle=0,width=\linewidth]{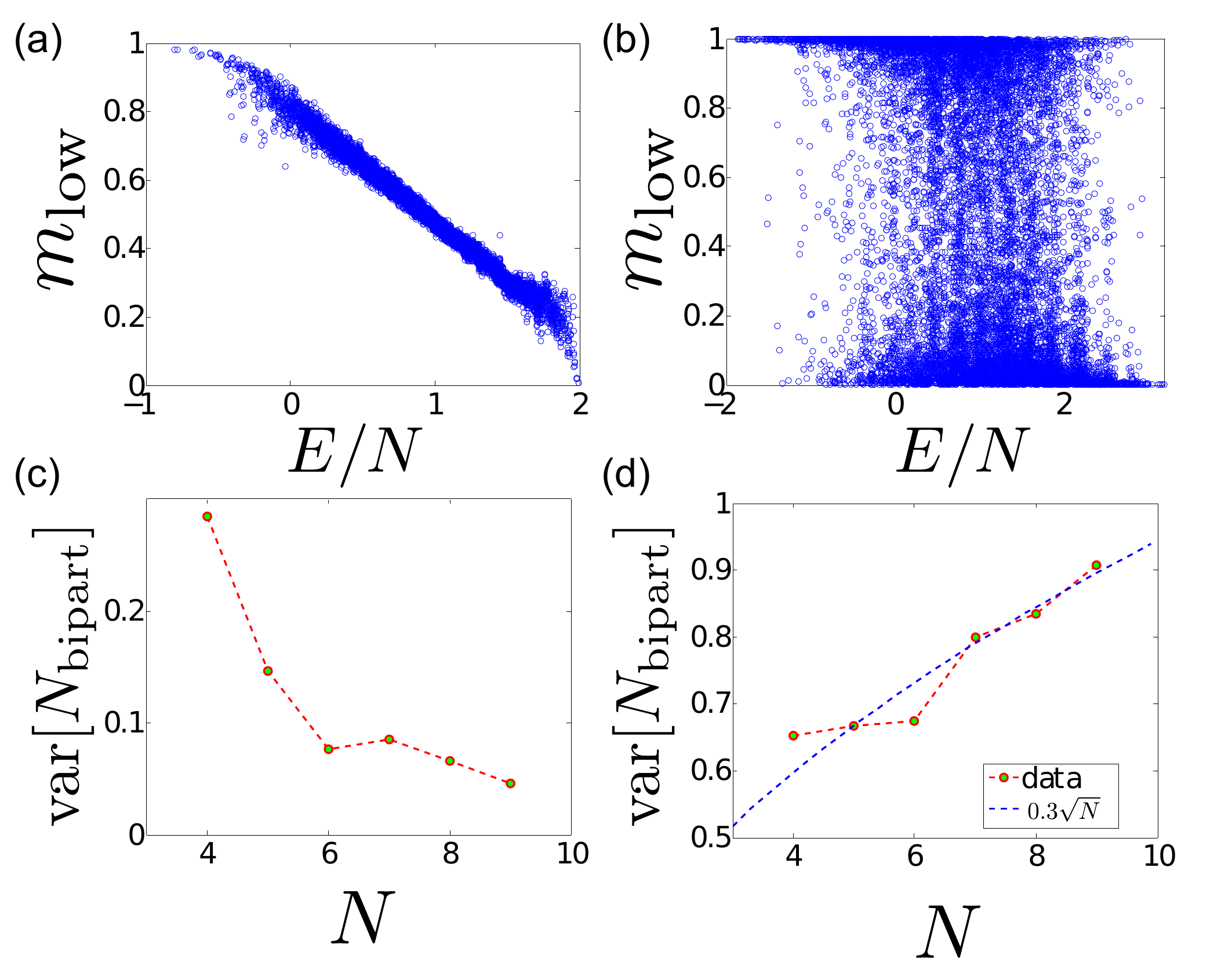}
 \caption{Eigenstate thermalization hypothesis and its violation in interacting Aubry-Andr\'e model. 
 (a) and (b) show the occupation number $m_{\rm low}$ of the lowest energy orbital. (c) and (d) show scaling of the fluctuation of the bipartite 
 ($\gamma = \frac{1}{2}$) occupation number versus the total particle number. The numerical results for this interacting case are consistent with the $\sqrt{N}$ scaling as we find for the non-interacting case. The interaction strength is fixed to be $V/t = 2$ here. (a) and (c) correspond to the delocalized phase where we choose $\lambda/t = 0.25$. (b) and (d) correspond to the localized phase with $\lambda/t = 2$.  }
\label{fig:1DIntAA} 
\end{figure}

For non-interacting fermions, the Hamiltonian is quadratic and can be written in a diagonal form, 
$$ 
H = \sum_\epsilon \epsilon \, n_\epsilon,  
$$ 
with $n_\epsilon$ the occupation number operator of the single-particle eigenstates.   
The many-body states in this basis are 
$$ 
|\Psi\rangle  = |m_1, \ldots, m_D \rangle, 
$$  
with $\langle \Psi| n_\epsilon |\Psi \rangle = m_\epsilon$ . The corresponding total energy is $E = \sum_\epsilon m_\epsilon \epsilon$. 
In the thermodynamic limit with the system size $D = L^d \to \infty $ and particle number $N\to \infty$, the energy $E$ is an extensive quantity and is almost independent of the occupation number of any particular orbital. For different many-body states near a certain given energy $E_0$, the occupation number $m_\epsilon$ can be treated as a random variable (see Fig.~\ref{fig:3DAnderson}), satisfying a distribution
\be 
P(m_\epsilon) = 
\left\{ 
\begin{array}{cc} 
 p_\epsilon & \text{for $m_\epsilon = 1$;}  \\   
1-p_\epsilon & \text{for $m_{\epsilon} = 0$. } 
\end{array}
\right. , 
\label{eq:Pmalpha} 
\ee 
 with $p_\epsilon = \frac{1}{e^{\beta (\epsilon-\mu) }+1}$, and $\beta$ and $\mu$ determined by the energy $E_0$ and the particle number $N$~\cite{StatMech}. This fermonic probability distribution is a statistical fact 
 (arising simply from the Pauli principle) 
 and does not rely on thermalization or ergodicity. For non-interacting fermions, we have strong fluctuations in $m_\epsilon$, characterized by its variance ${\rm var} [m_\epsilon] = \sqrt{ \overline{ m_\epsilon ^2 } - \overline{m_\epsilon} ^2 }= \sqrt{ p_\epsilon (1-p_\epsilon) }$. Therefore, the eigenstate thermalization hypothesis (ETH) is trivially violated for non-interacting fermions and the system is non-thermal 
 (as one would expect for a system with no inter-particle interactions). 
 In contrast, for an interacting system in a thermal phase, the fluctuation in $m_\epsilon$ is greatly suppressed and $m_\epsilon$ takes a value between $0$ and $1$ (see Fig.~\ref{fig:1DIntAA} (a)). With weak interactions, $m_\epsilon$ itself, instead of the probability $p_\epsilon$, obeys the Fermi-Dirac distribution and is equal to  $\frac{1}{e^{\beta (\epsilon-\mu) }+1}$. In this case ETH is respected.

\begin{figure}[htp!] 
\includegraphics[angle=0,width=\linewidth]{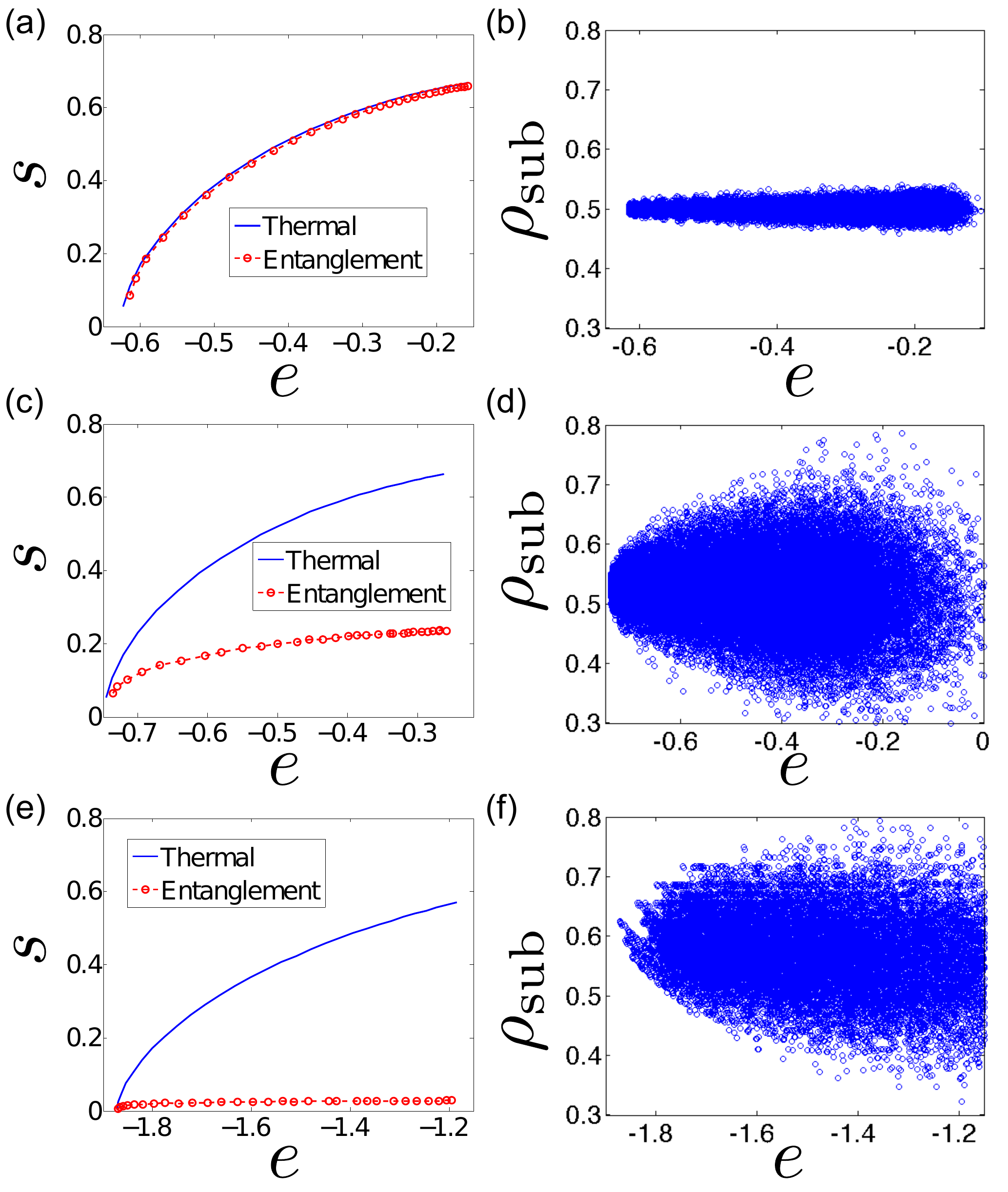}
 \caption{ETH violation in the many-body states of 1d Anderson model. (a), (c), and (e) show the thermal and entanglement entropy per lattice site ($s$) versus the energy density ($e$) for a subsystem.  The average density is fixed to be at half filling. (b), (d), (f) show the particle density in the subsystem ($\rho_{\rm sub})$. In this plot we choose a subsystem of $32$ sites within a global system with size $L = 2048$. The calculations are carried out using the stochastic sampling method illustrated in Sec.~\ref{subsec:sampling}, where the total particle number is not strictly fixed.  In (a,b), (c,d),and (e,f), the disorder strengths ($W/t$) are respectively, $0$, $2$, and $10$. For the clean case ($W/t=0$), the entanglement entropy is equal to the thermal entropy and the subsystem density does not exhibit strong fluctuations, 
 i.e. these observables look thermal despite the model being non-interacting.  
 For the disorder case, the entanglement entropy is smaller than the thermal entropy and the subsystem density shows strong fluctuations. 
 }
\label{fig:1dAndersonNonETH} 
\end{figure}

However, the local observables for a subsystem, average particle number for example, may look thermal and obey ETH even for non-interacting fermions (see Fig.~\ref{fig:1dAndersonNonETH}).  In this manuscript, we mainly consider non-interacting fermions with disorder,  and focus on ETH and its violation based on whether or not local observables are thermal. The notion of ETH for local observables was also previously used in studies of quantum thermalization~\cite{2011_Gogolin_ETH_PRL,2013_Ueda_ETH_PRE,2015_Yang_ETH,2015_Alba_ETH_PRB,2015_Magan_FreeFermion_arXiv,2016_Gluza_Gaussification_arXiv}. It is important to emphasize that if we instead study non-local observables, then non-interacting fermions are trivially non-thermal, essentially by definition.

\section{Diagnostics for localization and quantum nonergodicity} 
\label{sec:Diagnostics} 

To diagnose the localization transition, we calculate the infinite temperature entanglement entropy  by averaging over all eigenstates. 
We note here that although temperature itself is not a meaningful concept here (since there is no bath defining the temperature), the `infinite temperature' is a meaningful construct implying that all states are equally populated independent of their energies. 
In our calculation,  the average is carried out by randomly sampling the many-body spectra as the deterministic average is impossible due to the huge Hilbert space dimension, say for a system of size $L = 100$ at half filling. To characterize the thermal-to-nonthermal transition, we calculate the fluctuation of particle number in a subsystem, to be elaborated more in the following. With these two distinct diagnostics 
(i.e. entanglement entropy and number fluctuations), 
we 
are in a position to
distinguish the non-thermal transition from the localization transition. 

In this work, we investigate three model Hamiltonians exhibiting single-particle localization. 
One is the Anderson model describing fermions in random quenched disorder potentials with a Hamiltonian $H = H_0 + H_{\rm int}$~\cite{AndersonLocalization} 
\bea
&& H_0  = -t \sum_{\langle {\bf r}, {\bf r}'\rangle } \left( c_{\bf r} ^\dag c_{{\bf r}'} + {\rm H.c.} \right) + \sum_{\bf r} h_{\bf r}   c_{\bf r} ^\dag c_{\bf r} \nn \\ 
&& H_{\rm int} = V \sum_{\langle {\bf r}, {\bf r}' \rangle}  n_{\bf r} n_{{\bf r}'}, 
\label{eq:Anderson} 
\eea 
with $h_{\bf r}$ a random number drawn from a box distribution within $[-W/2, W/2]$, 
on a simple cubic lattice with periodic boundary conditions. 
(In this paper, we mostly study many-body properties in the noninteracting situation with $V=0$ for all our localization models-- we note that for fermions, even the noninteracting many-body problem is nontrivial with considerable correlations in the many-body product states.)
We study both one dimensional and three dimensional systems. The second model we study is a one dimensional generalized Aubry-Andr\'e (GAA) model describing fermions moving in an incommensurate potential, with a Hamiltonian $H = H_0 + H_{\rm int}$~\cite{sriramgaa,2015_Li_MBL_PRL}
\bea 
H_0 & = & -t \sum_{j=1}^{L} (c_j ^\dag c_{j+1} + H.c.) +
 2\lambda \frac{ \cos (  Q  j + \phi) }{1-\alpha \cos ( Q  j + \phi)  } n_j  
 \nonumber    \\
H_{\rm int} &=&  V \sum_j n_j n_{j+1}. 
\label{eq:GAA} 
\eea 
With $\alpha = 0$, this model reduces to the well-known Aubry-Andr\'e (AA) model~\cite{AA,azbel,harper55}. With $\alpha\neq 0$, the GAA model manifests a single-particle mobility edge~\cite{sriramgaa}. In this work $Q/2\pi$ is fixed to be the inverse golden ratio 
although in general any irrational number would work. 
The third model we study is an incommensurate non-interacting 1d lattice model with finite-range tunneling~\cite{2010_Biddle_PRL}
\bea 
&& H_0 \nn \\ 
&&  = -t \sum_{j \neq j'} e^{-p (|j-j'|-1)} c_j ^\dag c_{j'}  + 2 \lambda \sum_j   \cos (  Q  j + \phi)  n_j . 
\label{eq:HamNonLocal} 
\eea 
This model also has a single-particle mobility edge~\cite{2010_Biddle_PRL,biddleprb11}. 

\subsection{Entanglement and thermal entropy} 
\label{subsec:thermalvsent} 

Without interactions, the many-body eigenstates of fermions (see Appendix) 
are Slater-determinant product states (in order to satisfy the Pauli principle), whose entanglement entropy is fully determined by the two-point correlation function~\cite{2003_Peschel_EntEntropy} 
$$
{\rm G} _{ij} = \langle c_i ^\dag c_j \rangle. 
$$ 
Diagonalizing the correlation matrix ${\rm G}$ restricted to a local subsystem, we get eigenvalues $z_m \in [0, 1]$. The Von-Neumann entropy  is then simplified  for the non-interacting eigenstates to be 
\be 
S (l) = - \sum_m \left\{ (1- z_m) \log (1-z_m) + z_m \log z_m \right\}, 
\label{eq:ententropy} 
\ee 
with $l$ the number of lattice sites in the subsystem. 
The rank-$n$ R\'{e}yni entropy is given by 
\be 
S_n (l) = \frac{1}{1-n} \sum_m \log \left[ z_m ^n + (1-z_m) ^n \right] 
\ee 
In this work, we will mainly use Von-Neumann entropy of Eq.~\eqref{eq:ententropy}  for the entanglement entropy although our conclusions do not change if R\'{e}yni entropy is used.

If ETH is satisfied,  the subsystem is thermal and its local observables are described by the grand canonical ensemble. The thermal entropy of the subsystem then reads~\cite{StatMech} 
\be 
S_T  = \sum_m \left\{ \frac{\beta (\epsilon_m - \mu) } {1+e^{\beta (\epsilon_m -\mu) } } 
	+ \log \left[ 1+ e^{-\beta (\epsilon_m -\mu) } \right] 
	\right\}. 
\ee 
Here $\epsilon_m$ is the single-particle spectra of the subsystem, and $\beta$ and $\mu$ are the inverse temperature and  chemical potential of the thermal ensemble. 
{In the mathematical description of the thermal ensemble, the temperature is simply a parameter, while physically one can think that the subsystem is put in contact with an auxiliary external bath at a specific temperature. 
}
We use units such that the Boltzmann constant is unity.
At infinite temperature, the thermal entropy is related to the number of orbitals in the subsystem $l$ by 
$$ 
S_T = l \log 2. 
$$ 
This infinite temperature thermal entropy formula still holds in the presence of interactions.

It is worth noting that the distribution of many-body energies is typically strongly peaked at the middle of the spectra, $E_{\rm middle}$, regardless of localization or thermalization (see Appendix~\ref{sec：MBDOS}). 

To construct an eigenstate with energy equal to the infinite temperature energy, we can just pick an eigenstate at $E_{\rm middle}$. Alternatively, due to the peaked structure of many-body density-of-states (MBDOS), we can average over all eigenstates, since  
the main contribution to the average comes 
from the states very close to $E_{\rm middle}$. More practically for a large system, instead of using a deterministic average, we can perform a stochastic average by randomly sampling the many-body spectra with equal probability.  The equal-probability random sampling corresponds to an infinite temperature ensemble average
(in fact, this equal sampling of all states is the definition of the infinite temperature system). 
The stochastic method to sample eigenstates away from the spectra-center is to be discussed in Sec.~\ref{subsec:sampling}.

Taking an eigenstate with $E_{\rm middle}$, or equivalently averaging over the whole many-body spectra, we have ${\rm G}_{ij} = \frac{1}{2} \delta_{ij}$ in a thermal/ergodic phase. (To compare with the grand canonical ensemble, we should consider the total Hilbert space without fixing the total particle number.)  Then it follows that the entanglement entropy (Eq.~\eqref{eq:ententropy}) is strictly $l\log 2$. 
Considering a localization transition by varying the disorder strength, the ergodic phase with $S = l\log2$ is in general  stable upto a certain disorder strength $h_c$. Then there are two possibilities at $h_c$: (i) the entanglement entropy develops a discontinuous jump to an area law scaling, or (ii) it becomes smaller than $l\log2$, 
 i.e. sub-thermal. 
 But either of these two possibilities would imply that the entanglement entropy has a non-analytic behavior at $h_c$. For (i) $S$ itself is discontinuous; and for (ii) its first derivative is discontinuous. The latter is the most generic situation. In sections (~\ref{sec:AAtransition} and \ref{sec:GAAtransition}), we show that both scenarios could manifest in incommensurate lattice models.
For the one dimensional Anderson model, the entanglement entropy does not reflect the thermal entropy for any disorder strength since the system is always localized for any disorder strength (see Fig.~\ref{fig:1dAndersonNonETH}). 
  For an interacting system, there is yet a third possibility we cannot rule out, which is, (iii) the entanglement entropy does not reflect the thermal entropy even in a thermal phase. But this last possibility is unlikely.

\begin{figure*}[htp!] 
\includegraphics[angle=0,width=\linewidth]{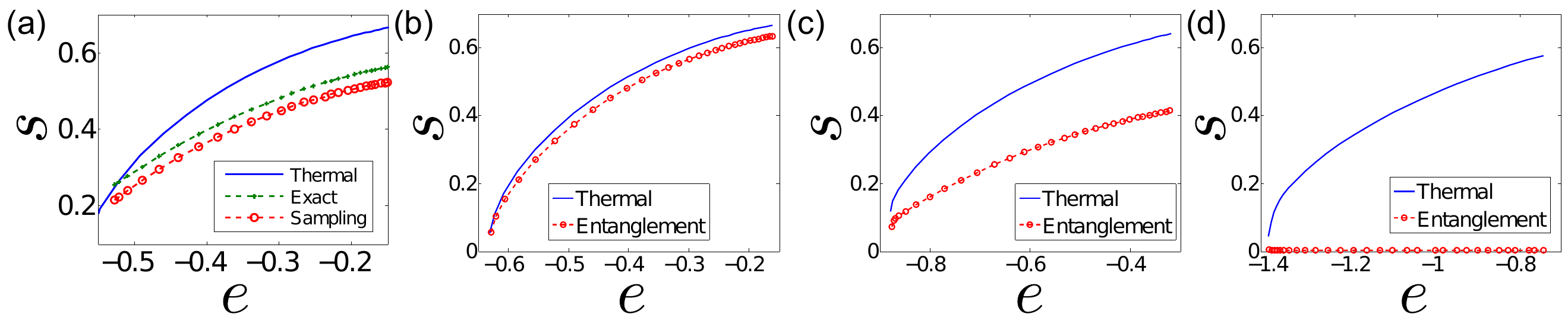}
 \caption{Thermal and entanglement entropy across the localization transition. Here we take the Aubry-Andr\'e model for illustration. We calculate the thermal and entanglement entropy per lattice site ($s$)  versus the energy density ($e$) for a subsystem. We use the stochastic method described in Sec.\ref{subsec:sampling}.  The average density is fixed to be at half filling. In (a), we choose a sub set of  four lattice sites ($L_{\rm sub} = 4$) within a system of size $L = 24$. For this small system, we calculate $s$ with both methods of using the exact spectra and the sampling method (see Sec.~\ref{subsec:sampling}). The difference between the results from the two methods is due to finite size effect and approaches zero as we increase the system size. We choose system sizes with $L_{\rm sub} = 128$ and $L = 2048$ in (b), (c), (d), which correspond to $\lambda/t = 0$ (delocalized phase), $\lambda/t =1$ (critical), and $\lambda/t = 2$ (localized phases), respectively.
}
\label{fig:ThermalVsEnt} 
\end{figure*}

From our numerical results shown in Fig.~\ref{fig:ThermalVsEnt}, it is confirmed that the entanglement entropy indeed reflects the thermal entropy in the delocalized phase, and both of them are extensive at finite-energy density. 
(Small deviations in the numerical results arise from finite size effects.)
In the localized phase, the thermal entropy remains extensive whereas the entanglement entropy becomes intensive. 
At the critical point ($\lambda =t $) of the Aubry-Andr\'e model, both entropies are extensive, but the entanglement entropy strongly deviates from the thermal value.

\subsection{Stochastic method to sample non-interacting eigenstates away from spectra center}  
\label{subsec:sampling}
As discussed in Sec.~\ref{subsec:thermalvsent}, the MBDOS is sharply peaked at the center of many-body energy spectra. It is thus unavoidable for  the non-interacting many-body eigenstates generated by equal-probability random sampling to fall into an energy region near the spectra center. Practically, this equal-probability sampling method cannot be used to generate eigenstates with a given energy away from the spectra center.  

Rigorously, we would like to sample the occupation of single-particle orbitals with hard constraints 
\bea 
\sum_{\epsilon} n_\epsilon = N \nn \\ 
\sum_{\epsilon} n_\epsilon \epsilon = E 
\label{eq:sampleconstraint}.  
\eea  
But it is difficult to implement such a  constrained sampling. 
As discussed in Sec.~\ref{sec:quadraticfluc}, from the constraint in Eq.~\eqref{eq:sampleconstraint}, $n_{\epsilon}$ satisfies a probability distribution (see Eq.~\ref{eq:Pmalpha}), which is purely a statistical consequence and does not rely on thermalization. We can thus sample the occupation of each single-particle orbital according to this distribution. In this way, the constraints are approximately satisfied. In Fig.~\ref{fig:ThermalVsEnt}(a), we explicitly show that this sampling method does reproduce the exact deterministic results. This sampling method has the added advantage of being adaptable to large systems  whereas the deterministic method is limited to very small systems. It is worth emphasizing that the sampling method does not rely on the system being thermal. It also works equally well for localized systems (see Fig.~\ref{fig:ThermalVsEnt}(d)).

\subsection{Scaling of particle number fluctuations in a subsystem} 
\label{sec:particlenumberscaling} 

In order to establish the `thermal' or `nonthermal' nature of the many-body system,
we consider fluctuations of average particle number in a local subsystem, 
\be 
N_{\rm sub} = \sum_{j = 1}^{\gamma D} \langle n_j \rangle .  
\ee 
Here $ n_j  =  c_j ^\dag c_j $, and $j$ runs over the sites in the subsystem with size $\gamma D$ ($\gamma < 1$). $D$ is the total number of lattice sites ($=L^d$ for a $d$-dimensional cubic lattice) in the whole system and $\gamma$ is the relative size of the subsystem (relative to the total lattice sites).  In the following we assume $\gamma \ll 1$ 
since we are trying to figure out if the rest of the system can thermalize the subsystem. 
At the same time $\gamma D$ should be taken as an extensive quantity, approaching infinity in the thermodynamic limit. 
Thus, in accordance with the standard limiting procedure in thermodynamics and statistical mechanics, both the system and the subsystem are becoming infinite in size with the constraint system much larger than the subsystem (i.e. $\gamma\ll 1$).

With random uncorrelated disorder, we can reasonably assume that the local densities,  $\langle n_j \rangle$ with different $j$s, are independent random variables. Then we know from the central limit theorem that $N_{\rm sub}$ satisfies the standard distribution,
\be 
P(N_{\rm sub} ) = \frac{1}{\sqrt{2\pi} \sigma_{\rm sub} } \exp \left[ -\frac{( N_{\rm sub} -\overline{N_{\rm sub}} )^2}{2 \sigma_{\rm sub} ^2} \right], 
\ee 
with $\overline{N_{\rm sub}} = \gamma N $. 
Here $\overline{ \ldots}$ means averaging over an energy shell.    
The fluctuation in $N _{\rm sub}$ is related to local density fluctuation by  
\bea 
&& \sigma_{\rm sub} ^2  = \gamma D \sigma_n ^2 \nn \\ 
&& \sigma_n ^2  =  \overline { \langle n_j \rangle ^2 } - \overline { \langle n_j \rangle} ^2 .  
\eea  
For non-interacting fermions, the density is determined by single-particle wave functions, $\varphi_\epsilon (j)$, 
as 
\be 
\langle n_j \rangle = \sum_{\epsilon }  m_{\epsilon} |\varphi_\epsilon (j) | ^2.
\label{eq:njwavefunc} 
\ee  
The fluctuation $\sigma_n$ is different for localized and extended wavefunctions. 

For extended states,  $\varphi_\epsilon (j)$ is a delocalized wavefunction, and the amplitudes $m_{\epsilon} | \varphi_\epsilon (j) |^2$ (with fixed $j$) behave as independent random variables with a standard deviation proportional to $D^{-1}$. It follows that 
$ 
\sigma_{\rm sub} ^2 \sim \rho = N/L^d,
$  
for an extended state, and the particle number fluctuation in a  local subsystem is thus an intensive quantity 
and does not diverge in the thermodynamic limit. 

For localized states, the quantity  $m_\epsilon |\varphi_\epsilon (j)| ^2$ (for different $\epsilon$ with fixed $j$) can no longer be treated as independent random variables due to the Pauli exclusion principle. Then the central limit theorem does not apply, and we need to work out the distribution of $\langle n_j\rangle$ more carefully. 
Consider a system with localization length $\xi$. In the summation to calculate $n_j$ (Eq.~\eqref{eq:njwavefunc}), there are approximately $\xi^d$ orbitals in the neighborhood of $j$ that contribute significantly. The probability of occupying $X$ out of $\xi^d$ number of orbitals is 
\be 
P( X) = 
\binom { \xi ^d}{X}  
\times \binom{L^d - \xi ^d}{N- X} 
/ 
\binom{L^d}{N}. 
\label{eq:PX} 
\ee 
The local density is related to $X$ approximately by $\langle n_j \rangle = X/\xi^d$. 
In a deep localized phase $\xi \approx 1$, then we have $P( \langle n_j\rangle = 0) = 1-\rho$ and $P(\langle n_j \rangle = 1) = \rho$, from which it follows $\sigma_n^2  = \rho - \rho^2$. Away from the deep localized phase, $\xi \gg 1$, we find by using Stirling approximation that 
\be 
P (\langle n_j \rangle ) \propto \exp \left\{ -\frac{\xi^d (1+ \xi^d/L^d) (\langle n_j \rangle -\rho) ^2 }{2\rho (1-\rho) } \right\}.  
\ee 
Then it follows that 
\be 
\sigma_n ^2 = \frac{\rho (1-\rho) } { (1+ \xi^d/L^d ) \xi^d}. 
\ee 
Since this equation is satisfied for both $\xi \approx 1$ and $\xi \gg 1$, we expect it to hold in the whole localized phase. Then the number fluctuation for the localized phase is obtained to be,  
\be 
\sigma_{\rm sub} ^2 = \frac{\gamma N (1-\rho) }{ (1+\xi^d/L^d) \xi^d }, 
\label{eq:Nsubscaling} 
\ee 
which is an {\it extensive} quantity, in sharp contrast to the extended states. The $\sqrt{N}$ scaling of $\sigma_{\rm sub}$ is confirmed in numerics (see Fig.~\ref{fig:3DAnderson} (f)). 
This extensive scaling of the fluctuation in the localized states immediately implies a failure of thermalization since the fluctuations diverge in the thermodynamic limit.

Although the above analysis assumes non-interacting fermions, we expect the resultant scaling to hold even in the presence of interactions, 
since the arguments leading to the scaling are general statistical arguments using only the applicability or not of the central limit theorem (and the Pauli principle which always applies to fermionic states).
Taking the local integral of motion description of the interacting many-body localized phase, say with conserved charges $q_j$~\cite{2013_Serbyn_PRL,2014_Huse_MBL_PRB,chandran2015,Ros2015420}, we have 
\be 
n_j = \sum_{j'}  {\rm Tr} [n_j \tilde{q}_{ j'} ] \tilde {q} _{j'} 
\ee 
with $\tilde {q} $ referring to the single-body component of the charge. 
Then assuming 
$${\rm Tr} [n_j \tilde{q}_{ j'} ]\sim \exp (-|j-j'| /\xi_{\rm int} ),$$ 
where $\xi_{\rm int}$ is an interaction decay length scale~\cite{2013_Serbyn_PRL,chandran2015,Ros2015420}, 
the same argument used for Eq.~\eqref{eq:PX}, would yield the same scaling as in Eq.~\eqref{eq:Nsubscaling}  for interacting systems. 
We have done exact diagonalization for the interacting Aubry-Andr\'e model (Eq.~\eqref{eq:GAA}), and our numerical results are consistent with the $\sqrt{N}$ scaling (see Fig.~\ref{fig:1DIntAA} (d)) in the localized phase.   
 

It is worth noting here that the result in Eq.~\eqref{eq:Nsubscaling} does assume one single localization length. If the system involves multiple localization length scales, the above argument can be easily generalized which still leads to the $\sqrt{N}$ scaling.  But the localization length in Eq.~\eqref{eq:Nsubscaling} will be replaced by certain averaged value that will depend on details of the many-body states, assuming that there are no diverging localization length scales.

\subsection{Experimental relevance} 

Experimentally, it is generally challenging to prepare a many-body excited state due to the dense energy spectra (Fig.~\ref{fig:MBDOS}). It is thus quite nontrivial to directly probe the particle-number fluctuations among eigenstates 
in the laboratory although it is a perfectly well-defined thermodynamic quantity. 
However, the fluctuation effects could  manifest in quantum quench dynamics measurements~\cite{1991_Deutsch_ETH_PRA,1994_Srednicki_ETH_PRE,2008_Rigol_thermalization_Nature}. We can prepare an ensemble of initial product states, e.g., 
$$ 
|\Psi_{ \{ n_1, n_2, \ldots n_{D} \} } \rangle = (c_1 ^\dag) ^{n_1} (c_2 ^\dag) ^{n_2} \ldots (c_{D} ^\dag )^{n_D } |{\rm vac} \rangle. 
$$ 
The energy of  such states is $E = \sum_{j} n_j h_j$.  Then we can monitor the long time dynamics of subsystem particle number of the product states. For a thermal system, the outcome (long-time average) will only depend on the energy $E$ and can be described by the thermal ensemble 
as dictated by ETH, 
whereas it will strongly depend on the initial configuration $\{n_1, n_2, \ldots n_D \}$ for a localized system
since the local memory is preserved in the nonthermal localized system. 
The variance of the observables, as we vary initial-state configurations, would reflect the fluctuations among the eigenstates. In this way, the scaling of particle number fluctuations can be probed in experiments. To experimentally probe energies away from the spectra center, one should consider preparing initial states according to the stochastic method described in Sec.~\ref{subsec:sampling}.

\section{Many-body transitions of non-interacting fermions without single-particle mobility edge} 
\label{sec:AAtransition} 

In this section, we study the transitions of many-body states of non-interacting fermions in the absence of single-particle mobility edge. 
There are many well-known examples of such noninteracting models, mostly in one dimension (e.g. 1d Anderson model where all states are localized for infinitesimal disorder or 1d Aubry-Andr\'e model where all states are extended or localized according to whether $t>\lambda$ or $\lambda>t$ in Eq.~\eqref{eq:GAA} with $\alpha=0$), and it may be important to emphasize here that the vast majority of the numerical MBL studies in the literature~\cite{huse2015many,2015_Altman_review} focus entirely on the interacting versions of these non-generic noninteracting models which have no mobility edges (i.e. where the single-particle spectra are always either completely extended or completely localized depending on the parameters of the corresponding noninteracting Hamiltonian such as disorder strength for the Anderson model or the strength of the incommensurate lattice potential in the Aubry-Andr\'e model).
We use the entanglement entropy scaling as a diagnostic tool to find the localization transition, and use the subsystem particle number fluctuation to find the quantum nonergodic transition (i.e. thermal to nonthermal, as the fluctuations become extensive in the nonthermal phase). 
To be concrete, we take the Aubry-Andr\'e model,  and we expect the results to generically apply for 
fermionic localization 
models without single-particle mobility edges.


\begin{figure}[htp] 
\includegraphics[angle=0,width=\linewidth]{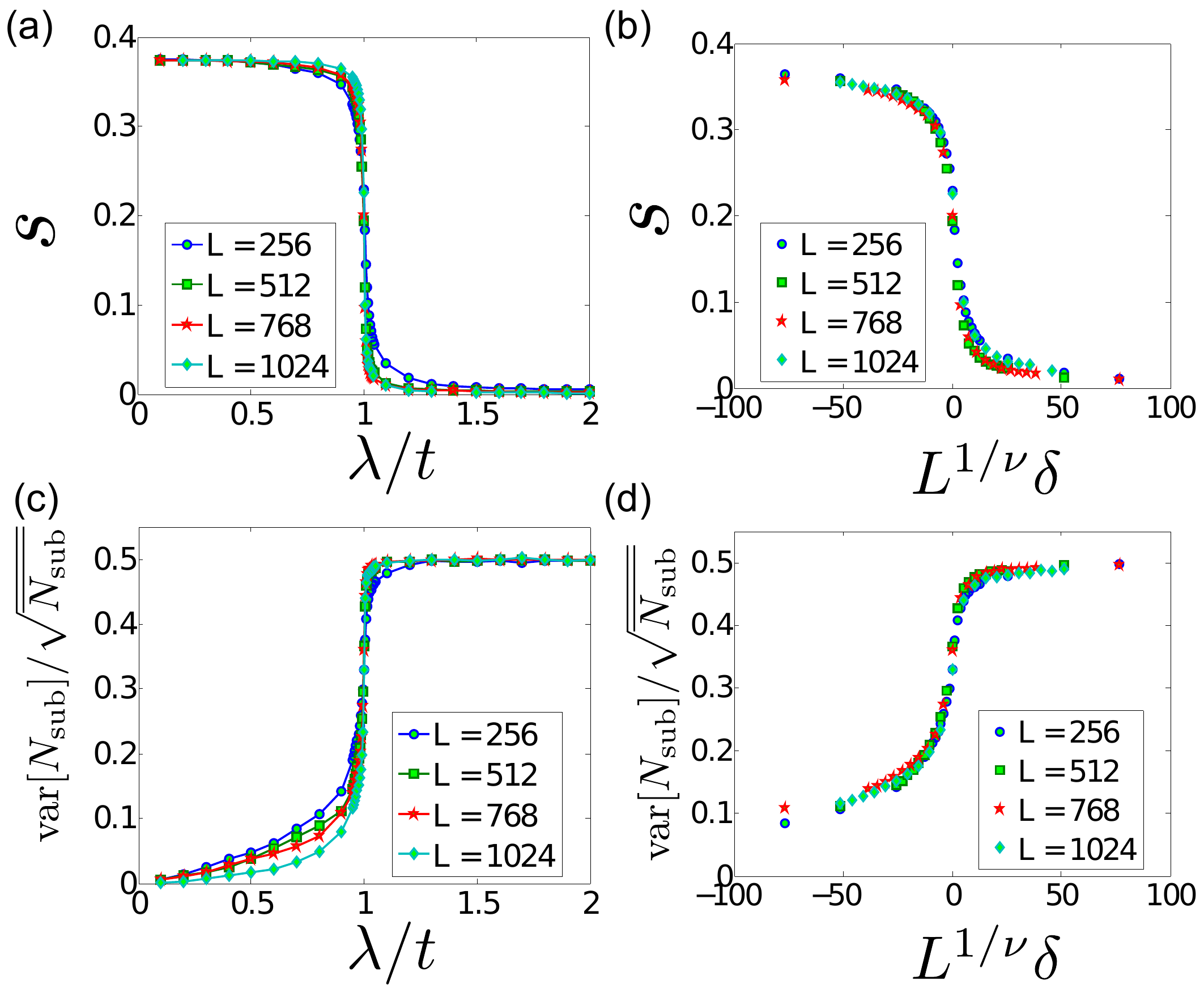}
 \caption{ 
 Phase transitions of non-interacting fermions in the Aubry-Andr\'e model at half filling. (a) and (b) show entanglement entropy density in one half of the system.  (c) and (d)  show the bipartite ($\gamma = \frac{1}{2}$)  particle number fluctuation. For both quantities, we average over all many-body eigenstates by stochastic sampling. (b) and (d) show the data collapse, where we take $\delta = (\lambda-\lambda_c) /t$, $\lambda_c = t$, and the scaling exponent $\nu = 1$. 
 From the system size dependence, we find the entanglement entropy is extensive (intensive) for $\lambda<t$ ($\lambda>t$). The bipartite  
 ($\gamma = \frac{1}{2}$) particle number fluctuation shows $\sqrt{N}$ scaling for $\lambda>t$, implying non-thermal behavior, whereas for $\lambda< t$, the fluctuation is greatly suppressed. Across the transition, both of the two quantities shown here, $s$ and ${\rm var} [\overline{N_{\rm sub }}] /\sqrt{N_{\rm sub} }$ develop a discontinuous jump as approaching the thermodynamic limit. In our calculation, we randomly sample $10^4$ ($10^5$) many-body states for (a,b) and (c,d), respectively. 
 }
\label{fig:1DAATransition} 
\end{figure}

In the Aubry-Andr\'e model (i.e., the tuning parameter $\alpha = 0$ in Eq.~\eqref{eq:GAA}), the single-particle wave functions have a transition at a critical incommensurate lattice strength $\lambda_c/t = 1$. All single-particle orbitals are localized (delocalized) for $\lambda$ above (below) the critical value $\lambda_c$. Across the single-particle localization transition, the normalized participation ratio vanishes continuously (see Sec.~\ref{sec:AAsingleparticle}),  
indicating vanishing conductivity in the localized phase. 
By constructing Slater determinants of single-particle orbitals 
(so as to form the appropriate noninteracting many-body fermionic states for the noninteracting system--see Appendix, 
we will now show that the many-particle non-interacting states exhibit a discontinuity in entanglement entropy, which is 
in sharp contrast with the single-particle case. 

\begin{figure}[htp] 
\includegraphics[angle=0,width=.8\linewidth]{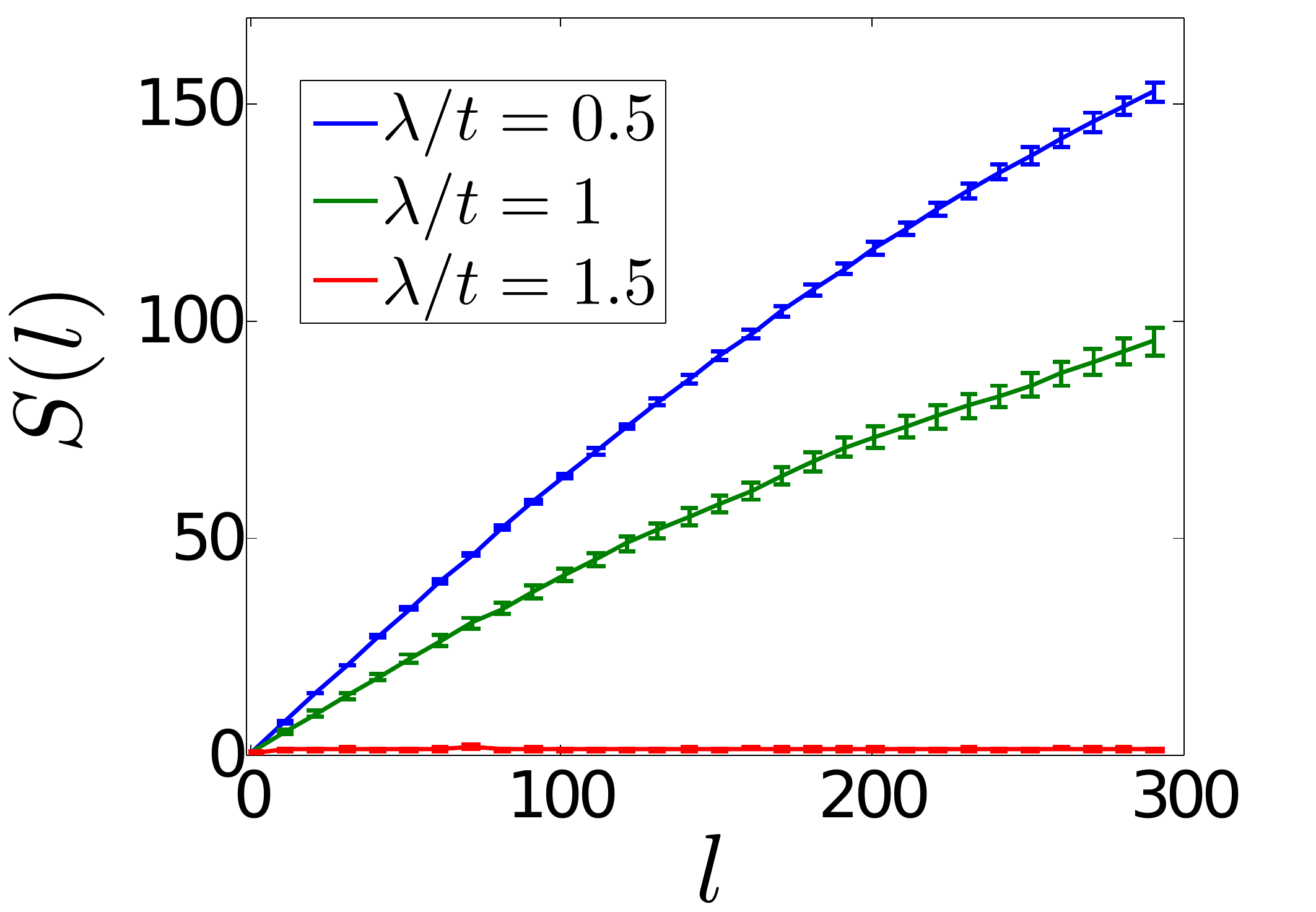}
 \caption{Entanglement scaling 
 as a function of the sub-system size $l$ 
 in localized, extended, and critical phases of the Aubry-Andr\'e model at half filling. 
 For $\lambda/t =0.5$, the system is in an extended phase, the entanglement entropy shows volume-law scaling, whereas for the localized phase with $\lambda/t=1.5$, the entanglement entropy shows area-law scaling. At the critical point $\lambda/t = 1$, the entanglement entropy exhibits volume-law scaling, with  slope lower than the fully extended phase.  We randomly average over $10^3$ many-body states here. We choose a system size  $L = 1000$. }
\label{fig:1DAAEntScaling} 
\end{figure}

In Fig.~\ref{fig:1DAATransition}, we show the calculated bipartite 
($\gamma = \frac{1}{2}$)  entanglement entropy and the particle number fluctuation. For weak incommensurate lattice strength $\lambda < t$, the entanglement entropy is extensive and the particle number fluctuation is systematically suppressed  as we increase the system size. The system is thus in an extended and thermal phase. Across the transition, i.e., for $\lambda>t$, the entanglement entropy becomes intensive and the particle number fluctuation is no longer suppressed, and has a $\sqrt{N}$ scaling. The large fluctuation indicates non-thermal behavior, whose experimental implications have been discussed in Sec.~\ref{sec:particlenumberscaling}.  In the thermodynamic limit both entanglement entropy and particle number fluctuation show a discontinuous jump at the transition point $\lambda/t = 1$. 
For finite size systems we find a clear crossing in the data for
both the entanglement entropy $S$ and the variance of the particle number directly
at the critical point, 
as can be seen in Fig.~\ref{fig:1DAATransition}. 
Thus we find that at the transition the
entanglement entropy is sub-thermal but still obeys the
volume law (and the variance in particle number scales
like $\sqrt{N}$) consistent with Ref.~\onlinecite{2014_Grover_MBL_arXiv}.  
We find a reasonable data collapse (Fig.~\ref{fig:1DAATransition}(b,d)) by taking a finite-size scaling form 
\be 
f (L ^{1/\nu_{\rm AA}}  (\lambda - \lambda_c) )
\ee 
with the length exponent $\nu _{\rm AA} = 1$ (see Section~\ref{sec:AAsingleparticle}). 
Right at the transition point $\lambda = \lambda_c = t$,  the entanglement entropy is still extensive but sub-thermal. Its volume-law scaling is explicitly shown in Fig.~\ref{fig:1DAAEntScaling}. At the same time, the particle number fluctuation exhibits $\sqrt{N}$ scaling. The system at this  critical point is thus extended but non-thermal, forming a fine-tuned nonergodic metal.  It is important to note that AA and GAA models do not satisfy the Chayes-Chayes-Fisher-Spencer (CCFS) bound~\cite{1986_CCFSBound_PRL} (see Sec.~\ref{sec:AAsingleparticle}) and therefore 
the AA critical exponents 
are not expected to be restricted by a mean theorem~\cite{2015_Chandran_Laumann_arXiv}. 

It is worth noting here that we expect the 
parameter-tuned
direct transition (as in Fig.~\ref{fig:1DAATransition}) 
from the thermal and extended phase  to the non-thermal and localized phase 
to be generic only in the absence of a mobility edge, 
(i.e. when all single-particle states in the system are extended or localized depending on parameter values). 
In the next section, we will show that this direct transition splits into two successive transitions in the generalized Aubry-Andr\'e model (with $\alpha \neq 0$ in Eq.~\eqref{eq:GAA}) 
because of the existence of a single-particle mobility edge.

\begin{figure}[htp] 
\includegraphics[angle=0,width=.8\linewidth]{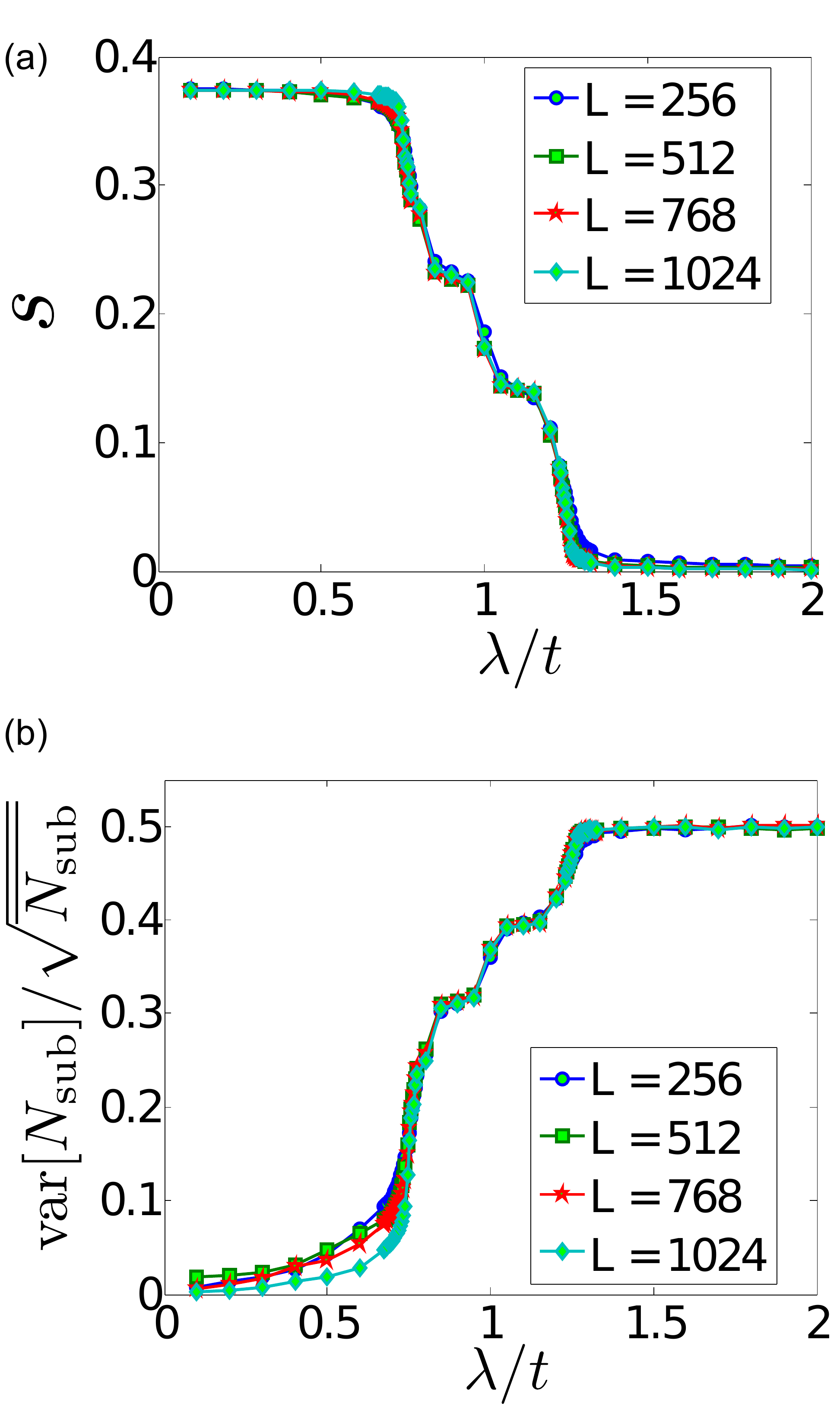}
 \caption{Phase transitions of non-interacting fermions in the GAA model (Eq.~\eqref{eq:GAA}) at half filling. (a) shows entanglement entropy density in one half of the system.  (b) shows the bipartite 
 ($\gamma = \frac{1}{2}$) 
  particle number fluctuation.  The tuning parameter $\alpha$ is fixed to be $-0.2$. We find the extensive-to-intensive transition for entanglement entropy is located at $\lambda_{L} \approx 1.3 t$, 
 and  the thermal-to-nonthermal transition is located at $\lambda_{T} \approx 0.75t$, i.e. the bipartite particle number fluctuation is strong (suppressed) above (below) $\lambda_{T}$. As we increase the system size, the two quantities shown here exhibit singular behavior at $\lambda_{T}$ and $\lambda_{L}$. The step-like structures in between root in the spectra gaps of the model. In our calculation, we randomly sample $10^4$ ($10^5$) many-body states for (a) and (b), respectively.    }
\label{fig:1DGAATransition} 
\end{figure} 

\begin{figure}[htp] 
\includegraphics[angle=0,width=.8\linewidth]{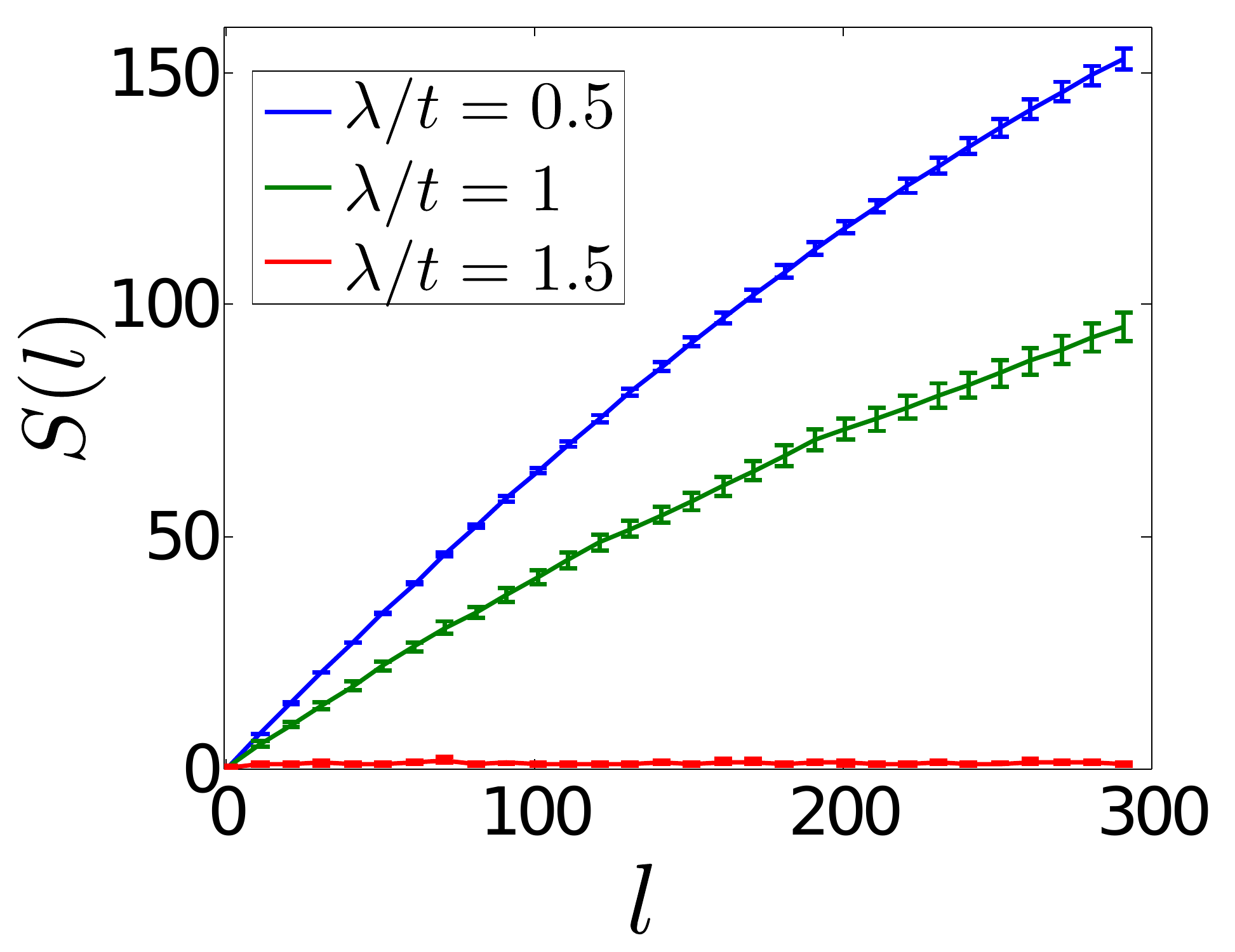}
 \caption{Entanglement scaling as a function of the sub-system size $l$ in localized, extended, and partially-extended phases of the generalized Aubry-Andr\'e model (Eq.~\eqref{eq:GAA}) at half filling. Like the Aubry-Andr\'e model ($\alpha=0$), the entanglement entropy shows volume law scaling for $\lambda/t = 0.5$ and area law for $\lambda/t = 1.5$. For $\lambda/t = 1$, the GAA model is in a partially-extended phase where the entanglement entropy shows volume law scaling. 
The subthermal partially extended states  are composed of both localized and delocalized single-particle orbitals. 
We randomly average over $10^3$ many-body states here.  We choose a system size  $L = 1000$.}
\label{fig:1DGAAEntScaling} 
\end{figure} 

\begin{figure}[htp] 
\includegraphics[angle=0,width=\linewidth]{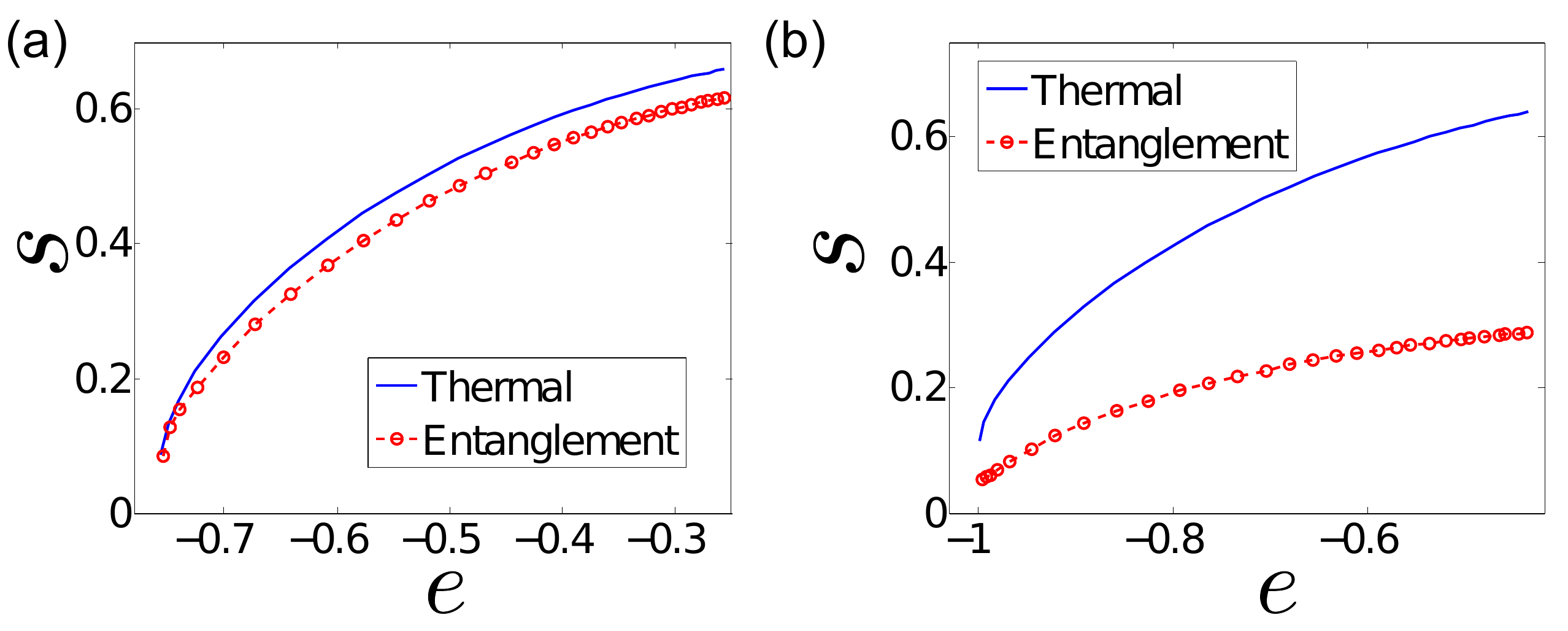}
 \caption{Thermal and entanglement entropies 
 as a function of the energy density $e$
 for the generalized Aubry-Andr\'e model (Eq.~\eqref{eq:GAA}) at half filling. In this plot we take the tuning parameter $\alpha = -0.2$ (Eq.~\eqref{eq:GAA}).  We calculate the thermal and entanglement entropy per lattice site ($s$)  versus the energy density ($e$) for a subsystem. The average density is fixed to be at half filling. We choose a sub set of $L_{\rm sub}$ lattice sites ($L_{\rm sub} = 128$) within a system of size $L = 2048$.   (a) and (b) correspond to $\lambda/t = 0.5$ (fully extended phase), and $\lambda/t =1$ (partially extended phase), respectively. In the fully extended phase shown in (a), the entanglement entropy is approximately equal to the thermal entropy. We expect the deviation is due to finite size effects. In the partially extended phase shown in (b), the entanglement entropy strongly deviates from the thermal entropy, the reason being partially extended states  are composed of both localized and delocalized single-particle orbitals. 
}
\label{fig:ThermalVsEnt_GAA} 
\end{figure}

\section{Many-body spectra with a single-particle mobility edge} 
\label{sec:MEtransition} 

In this section, we study the transitions of many-body states of non-interacting fermions in the presence of single-particle mobility edge. Due to the existence of a single-particle mobility edge,  besides putting all particles in localized orbitals or delocalized ones, there is yet a third possibility to construct non-interacting many-body states, which is to put some particles in the localized orbitals and others in the delocalized ones, 
so that the Slater determinant describing the many-body state at a particular many-body energy consists of both extended and localized single-particle orbitals-- this particular possibility does not exist in models without a single-particle mobility edge since the whole single-particle spectrum in this case must consist entirely of either all localized or all extended states.
Such states have been dubbed ``partially-extended states"~\cite{2015_Li_MBL_PRL}
 but they could equally well be called ``partially-localized" states. 
Now we investigate the entanglement scaling and  particle number fluctuations in 
models with single-particle mobility edges, in particular, 
a one dimensional incommensurate lattice  model (Eq.~\eqref{eq:GAA}) and a three dimensional random disorder model (Eq.~\eqref{eq:Anderson}), where we find unique features of the partially-extended phase distinct  from either the fully extended or the fully localized phase.

\subsection{Many-body transitions in  the  incommensurate lattice models with mobility edge} 
\label{sec:GAAtransition} 

Unlike the Aubry-Andr\'e limit (or the 1d Anderson model), the generalized incommensurate lattice model with finite $\alpha$ (Eq.~\eqref{eq:GAA}) could manifest a single-particle mobility edge whose analytic form has been worked out by constructing a spectral duality~\cite{sriramgaa}.  
The 1d GAA model is thus akin to the 3d Anderson model, with the single-particle energy spectra being sharply divided by a critical energy (i.e. mobility edge) into localized and extended states. 

In Fig.~\ref{fig:1DGAATransition}, we show the bipartite entanglement entropy and the particle-number fluctuation for the GAA model. 
The GAA model manifests two transitions---a thermal-to-nonthermal transition at $\lambda_{\rm T}$ and  a localization transition at $\lambda_{\rm L}$ with $\lambda_{\rm T} < \lambda_{\rm L}$.  Below $\lambda_{\rm T}$, the entanglement entropy is extensive and reflects the thermal entropy (see Fig.~\ref{fig:ThermalVsEnt_GAA} (a)). The entanglement entropy remains a constant as we increase $\lambda$ in the thermal phase. The particle number fluctuation (normalized by $\sqrt{N}$) vanishes in the thermodynamic limit. For  $\lambda> \lambda_{\rm T}$, the entanglement entropy remains extensive, but no longer reflects the thermal entropy (see Fig.~\ref{fig:ThermalVsEnt_GAA} (b)), which means the system is non-thermal. Consistently, the particle number fluctuation develops the non-thermal $\sqrt{N}$ scaling. The system is in a non-thermal extended phase in the parameter regime $\lambda \in (\lambda_{\rm T}, \lambda_{\rm L})$. Further increasing the incommensurate lattice strength, the system becomes fully localized when $\lambda > \lambda_{\rm L}$. In this parameter regime, the entanglement entropy becomes intensive and completely fails to track the thermal entropy. The particle number fluctuations then saturate to  a constant in this GAA model.

At the transition points $\lambda_{\rm T}$ and $\lambda_{\rm L}$,  the two examined quantities (shown in Fig.~\ref{fig:1DGAATransition})  are non-analytic in the thermodynamic limit, in the sense that their first derivatives are discontinuous. 
This implies that the intermediate non-thermal extended phase is indeed a well-defined unique phase.  For non-interacting fermions, the emergence of this intermediate phase originates from the existence of partially extended states, 
in the full many-body spectra by virtue of the mobility edge in the single-particle spectra.  The discussion and the results of this subsection are entirely for the noninteracting many-body GAA spectra where partially extended Slater determinant states are manifestly present.  What happens when inter-particle interactions are turned on (and the many-body spectra can no longer be constructed simply from the single-particle orbitals)?
For interacting fermions, the concept of partially-extended states does not carry over. But we do expect the intermediate non-thermal extended phase
to survive (since there is no obvious reason for this phase to disappear in the presence of a weak interaction)~\cite{2015_Li_MBL_PRL,modak2015many}. 
This is particularly true since this intermediate nonthermal extended phase
 is well defined by the non-analytic behavior of the entanglement entropy.  
 We expect this intermediate many-body phase to be generically present for interacting fermions whenever the corresponding noninteracting model has a mobility edge~\cite{2015_Li_MBL_PRL}. 

\begin{figure}[htp] 
\includegraphics[angle=0,width=.7\linewidth]{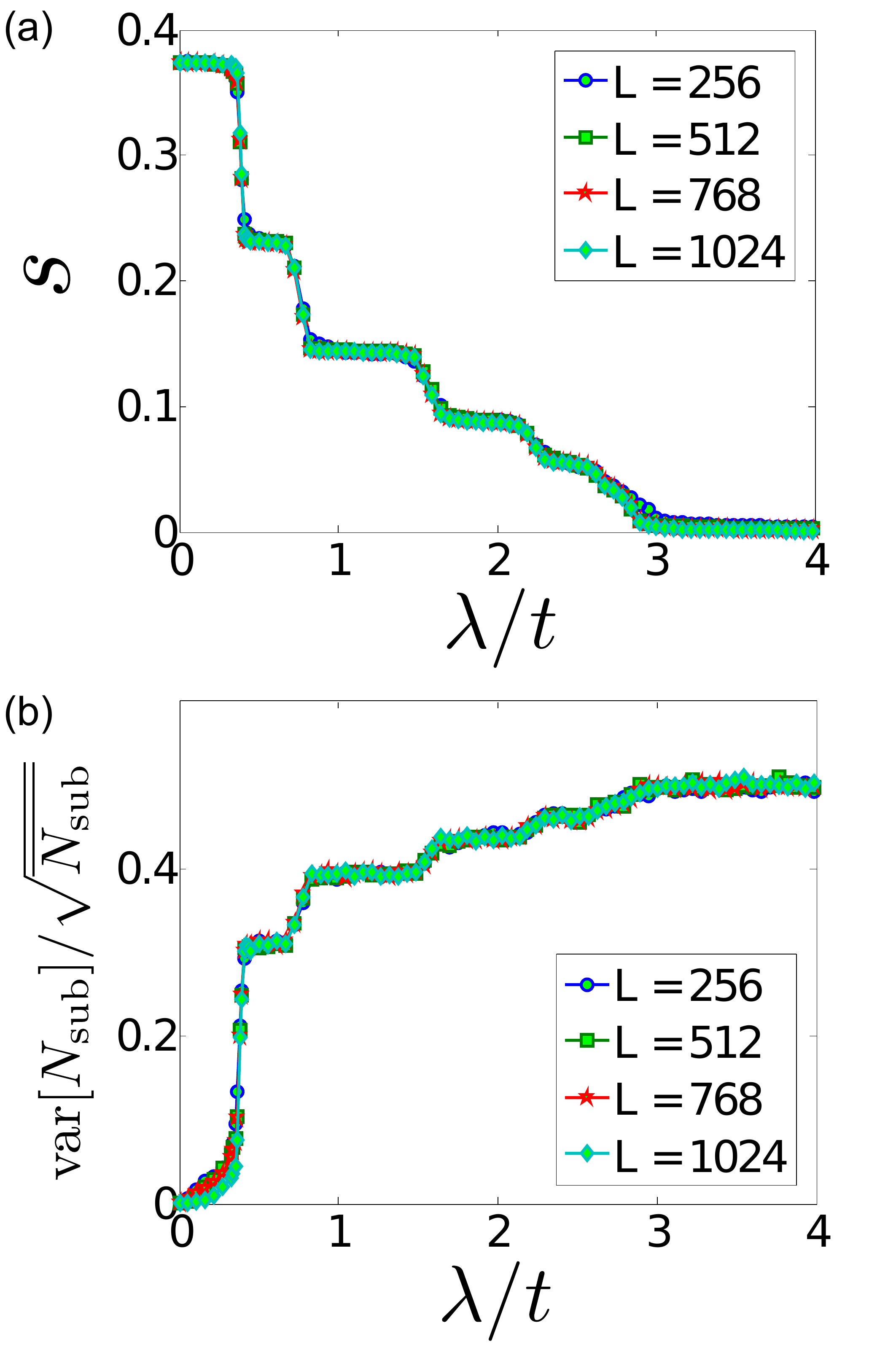}
 \caption{Phase transition of non-interacting fermions in the 1D nonlocal tunneling model at half filling (Eq.~\eqref{eq:HamNonLocal}). (a) shows entanglement entropy density in one half of the system.  (b) shows the bipartite ($\gamma = \frac{1}{2}$) particle number fluctuation. For this model, we find the extensive-to-intensive transition for entanglement entropy locates at $\lambda_{L} \approx 3 t$. The thermal-to-non-thermal transition locates  at $\lambda_{T} \approx 0.3 t$. In our calculation, we randomly sample $10^4$ ($10^5$) many-body states for (a) and (b), respectively. The coefficient $p$ controlling the non-locality is fixed to be $1$ here.  } 
\label{fig:1DNonLocalTunnel} 
\end{figure}

We have also examined the incommensurate lattice model with a non-local tunneling term in the Hamiltonian (Eq.~\eqref{eq:HamNonLocal}). This model, similar to the GAA model, also has a single-particle mobility edge, 
which can be obtained from an energy-dependent 
duality transformation~\cite{2010_Biddle_PRL}. From Fig.~\ref{fig:1DNonLocalTunnel}, the non-thermal and localization behavior of this model is qualitatively similar to the local GAA model described above, 
as is expected since both models have single-particle mobility edges.

\begin{figure}[htp] 
\includegraphics[angle=0,width=\linewidth]{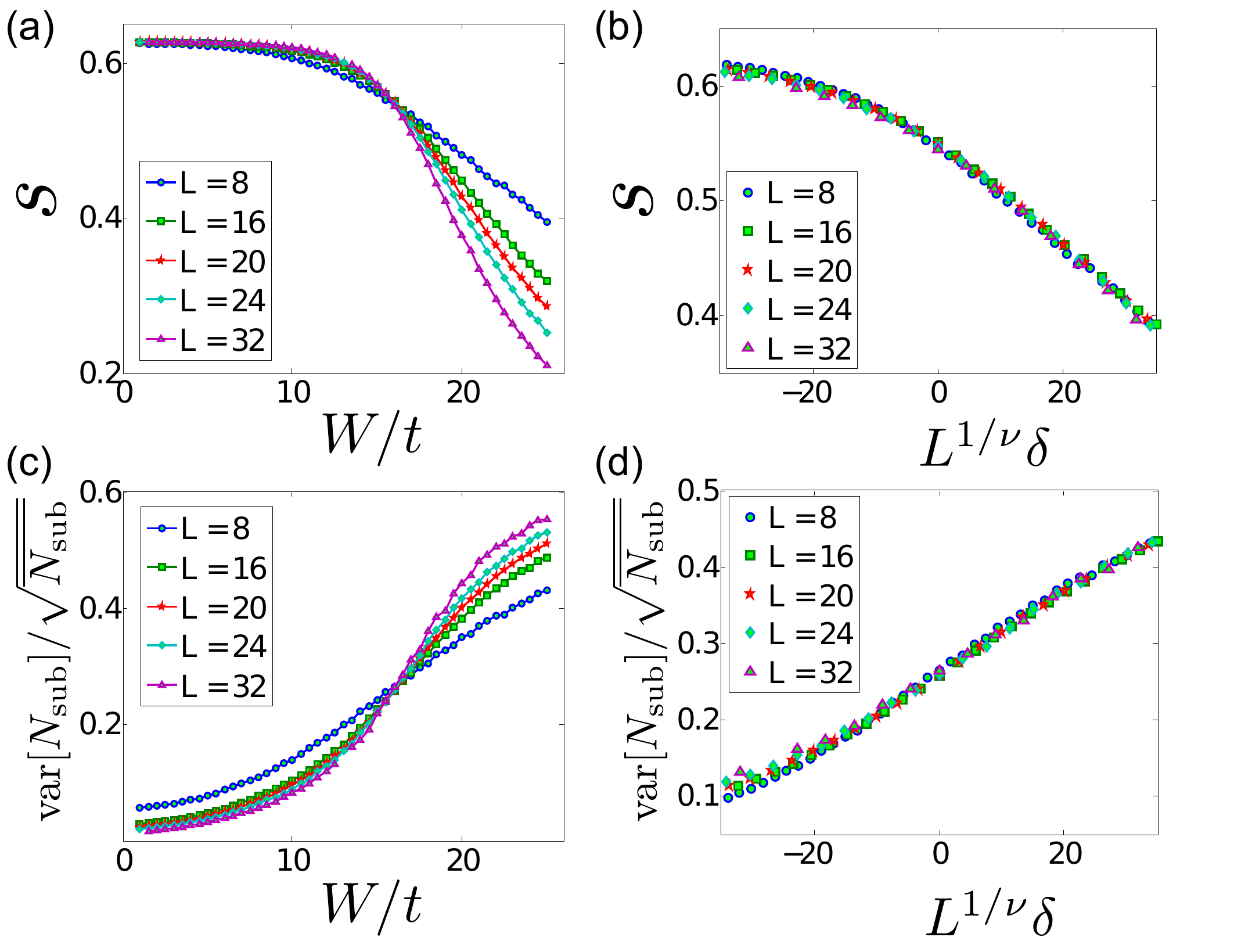}
 \caption{Phase transition of non-interacting fermions in the 3d  Anderson model at half filling. 
 We take a 3d system of size $L^3$ and calculate entanglement entropy per site [shown in (a) and (b)] and particle number fluctuation [shown in (c) and (d)], in  a subsystem (with size $(L/2)^3$). (b) and (d) show the data collapse, where we take $\delta = (W-W_c)/t$, $W_c/t = 16$,  and scaling exponent $\nu = 1.57$. For entanglement entropy, we randomly sample 
 $10^2$, and $10^3$ many-body states, respectively for large ($L = 24, 32$) and small  ($L= 8, 16, 20$) systems. For the particle number fluctuation we sample $10^4$ and $10^3$ states, for small and large systems. We average over $100$ ($10$) disorder realizations for  $L = 8$, $16$, and $20$  (for $L = 24$ and $L = 32$). With finite disorder strength, the particle number fluctuation saturates to a  finite value as we increase the system size. 
}
\label{fig:3DAndersonTransition} 
\end{figure}

\subsection{Many-body transitions  in the three dimensional Anderson model} 
In addition to the 1d incommensurate lattice models discussed above, we also investigate the many-body effects of the single-particle mobility edge  in the three dimensional Anderson model (Fig.~\ref{fig:3DAndersonTransition}) where the presence of mobility edge is a rule rather than an exception. 
Although both 1d GAA and 3d Anderson models have single-particle mobility edges (and hence they both manifest the intermediate nonergodic delocalized many-body phase in addition to the usual ergodic extended and nonergodic localized phases in their noninteracting many-body spectra), there is a significant difference between them.
The difference from the GAA model here is that the localized single-particle orbitals immediately kick in 
(existing for all disorder strength)
as we turn on the disorder potential, whereas in GAA model, the localized single-particle orbitals show up at some nonzero incommensurate lattice strength. Consequently, many-body states in the 3d Anderson model exhibit non-thermal behavior even for infinitesimal disorder, 
since some localized orbitals contributing to the many-body wavefunctions exist already for infinitesimal disorder (specifically at the single-particle band edges). 
This is consistent with our numerical results showing strong ($\sqrt{N}$ scaling) fluctuations in the subsystem particle number immediately after we turn on the disorder (see Fig.~\ref{fig:3DAndersonTransition} (c)). 
Thus, the many-body states in the noninteracting 3d Anderson model are generically nonergodic or nonthermal in the presence of any disorder (because of the invariable presence of the localized orbitals in the many-body  wavefunction) independent of whether the spectrum is extended or localized in the sense of the entanglement entropy (volume or area law). 
We remark that it is possible to choose a particular variant of the GAA model~\cite{sriramgaa} that has this same feature (as a function of $\lambda$) as the 3d Anderson model.
 The only situation where the 3d Anderson model allows for a thermal extended state spectrum is the trivial case (a set of measure zero) with no disorder, which is in sharp contrast with the 1d GAA model where the ergodic extended spectrum is generically present (in addition to the localized nonergodic and extended nonergodic states as well).

As shown in Fig.~\ref{fig:3DAndersonTransition}, the entanglement entropy of many-body states 
shows an extensive-intensive transition 
in the 3d Anderson model 
at the critical disorder strength $W_{L} \approx 16 \pm 0.5$, above which all of the single-particle orbitals are localized. 
As we are considering infinite temperature (i.e. equal-weight average over all many-body energy states), due to the shape of the mobility edge in energy and disorder strength~\cite{2006_Weisse_KPM_RMP}, we expect that the many-body (non-interacting) localization transition will occur when the single-particle problem localizes at the center of the band ($ W_L \approx 16.5$~\cite{1999_Slevin_Ohtsuki_PRL}), which is in agreement with our numerics,  
for the intensive-extensive transition in the entanglement entropy indicating that the transition in the entanglement entropy coincides with the ground state single-particle localization transition. 
Correspondingly, the particle number fluctuation shows qualitative difference in system size dependence below and above the critical disorder strength $W_{L}$. 
In both quantities we find a clear 
quantum critical 
crossing 
behavior 
for various system sizes. 
Using a finite size scaling form 
\be 
f (L ^{1/\nu_{\rm And} }  (W - W_L) ),  
\ee 
with the localization length critical scaling exponent $\nu _{\rm And}$ set to be $1.57$~\cite{1999_Slevin_Ohtsuki_PRL} (which is the known value for the single particle localization length exponent in the 3d Anderson model), we see data collapse for both particle number fluctuations and entanglement entropy density (Fig.~\ref{fig:3DAndersonTransition}(b,d)). 
This strongly suggests discontinuous jumps for both quantities in the thermodynamic limit for 
$W = W_L$. 
This  transition separates a non-ergodic extended phase ($W < W_L$) and a localized phase, 
and our results are consistent with a  ``mean" theorem~\cite{2015_Chandran_Laumann_arXiv}. 
This behavior also makes the 3d Anderson model different from the 1D GAA model. 
In particular, the 3d Anderson model seems to have just one critical disorder strength ($W_L$) (in contrast to two critical parameters  $\lambda_L$ and $\lambda_T$ of the GAA model studied in the last section) as there is just one single transition from a delocalized (volume law entanglement entropy) to a localized (area law entanglement entropy) with both phases having extensive nonthermal particle number fluctuations in the 3d Anderson model.  To be very precise, we can consider the lower critical disorder (i.e. $W_T$)  to be located at $W_T=0$ where the system makes a transition from the (zero-disorder) extended thermal state (where the particle number fluctuation is precisely zero) to the extended nonthermal state with $\sqrt{N}$ fluctuations in particle number at any finite disorder.  With this identification ($W_T=0$), the 3d Anderson model and the 1d GAA model become similar in their behaviors.

Due to heavy computational cost, for the 3d model even in the absence of any interaction, the largest system size we can afford to study  is a linear system size $L = 32$ 
(even for the noninteracting problem), 
which is much smaller than the largest $L$ we have used for the 1d GAA model. 
(This is simply because of the $L^d$ dependence of the Hilbert space dimensions which makes a linear system size of $L$ in three dimensions roughly equivalent to a linear system size of $L^3$ in one dimension.) 
In this regard, the 1d GAA model is an optimal platform to investigate mobility edge physics 
with respect to many body localization. However we emphasize here that the many-body 
localization transitions 
of non-interacting fermions in the 1d GAA and the 3d Anderson models appear to be  qualitatively different as discussed above. 
Whether these two models with single-particle mobility edges behave qualitatively similarly in the presence of interactions remains an interesting and important open question at this time since it is not possible to do meaningful MBL simulations of the 3d interacting Anderson model because of computational limitations. 
Based on our identification of $W_L$ and $W_T$ ($=0$) in the Anderson model, which correspond qualitatively to the critical parameters $\lambda_L$ and $\lambda_T$ in the GAA model, we speculate that the key feature of the extended nonergodic many-body phase (existing in the noninteracting 3d Anderson model for $W_T=0<W<W_L$) carries over to the interacting 3d Anderson model (as in the 1d GAA model) because we see no reason for this phase to disappear the moment the interaction is turned on~\cite{2015_Li_MBL_PRL,modak2015many}.  Only large scale future numerical work beyond the scope of the current work can settle this issue.

Regarding experimental relevance, it is important that particle number fluctuations exhibit similar critical behavior as the entanglement entropy in the 3d Anderson model.  Measurement of entanglement entropy for large systems is extremely challenging (if not impossible) in the laboratory, whereas particle fluctuations can be extracted from quench dynamics in a straightforward way as discussed in Sec.~\ref{sec:particlenumberscaling}.  Our results suggest that investigating quench dynamics 
experimentally to measure the number fluctuations 
may help probe critical properties of the 3d Anderson localization.

\subsection{Quantum nonergodic nature of the partially-extended phase} 
\label{sec:NonErgodicPEP} 
As we discuss above in the context of 1d GAA and 3d Anderson models, although partially-extended states have extensive entanglement entropy,  they are nonergodic in nature  
as manifested in the particle number fluctuations growing monotonically with the system size (and diverging in the thermodynamic limit).
We discuss several nonergodic aspects of the partially extended states below.

First, the many-body normalized participation ratio (NPR)~\cite{vadim}, which measures how the many-body wave function spreads over the Fock space, vanishes 
for the partially-extended many-body phase (indicating nominally a non-conducting behavior), 
as shown analytically in Ref.~\onlinecite{2015_Li_MBL_PRL}. For an ergodic system eigenstate wave functions are expected to approximately equally explore the Fock space, giving rise to  a finite NPR. For non-ergodic systems, the wave functions are not able to spread over the whole Hilbert space
(being localized in particular isolated regimes of the Fock space), 
for which the NPR vanishes. Taking a many-body state composed entirely of only localized single-particle orbitals, the NPR vanishes following a finite-size scaling form 
$ 
\eta \propto 1/ V_H , 
$ 
as we approach the thermodynamic limit.  For a partially-extended state, the NPR vanishes according to a different scaling formula 
$$
\eta \propto 1/V_H ^{\zeta},  
$$ 
with the exponent $\zeta \in (0, 1)$.  This behavior of partially extended states is similar to the non-ergodic extended phase studied in the context of single-particle localization in  Bethe lattices~\cite{2012_Biroli_arXiv,2014_Luca_BetheLattice_PRL}.  
Vanishing of NPR along with the volume law entanglement entropy indicate that the partially-extended-states cannot be considered either purely `metallic' or purely `localized' in the usual sense.  They are akin to `critical' wavefunctions in the ordinary ground state localization phenomena.

Second, the entanglement entropy of partially extended states does not reflect the thermal entropy. For an ergodic system, the entanglement entropy is expected to reflect the thermal entropy. We have shown for  the GAA model that the entanglement entropy in the fully extended phase is actually equal to the thermal entropy (Fig.~\ref{fig:ThermalVsEnt_GAA}).  For the partially extended phase, although the entanglement entropy obeys the volume law, it strongly deviates from the thermal entropy, again indicating its nonthermal and nonergodic character.

Third, the partially extended phase does not respect ETH  in a sense that its local observables exhibit strong non-thermal fluctuations. For a system respecting ETH, the fluctuations of local observables among nearby eigenstates are suppressed. In the partially extended phase, the local observables exhibit strong fluctuations (with $\sqrt{N}$ scaling), as we have shown by studying the local particle number for the GAA (Eq.~\eqref{eq:GAA}), 
the long-range hopping (Eq.~\eqref{eq:HamNonLocal}),  and 3d Anderson models (Eq.~\eqref{eq:Anderson}).  These fluctuations originate from localized orbitals composing the partially extended many-body states. 

We would like to mention here that the nonergodic metallic behavior has also been discussed in other contexts~\cite{1997_Altshuler_PRL,2012_Biroli_arXiv,2014_Luca_BetheLattice_PRL}. But its physical origin is very different from our case.

\begin{figure}[htp] 
\includegraphics[angle=0,width=\linewidth]{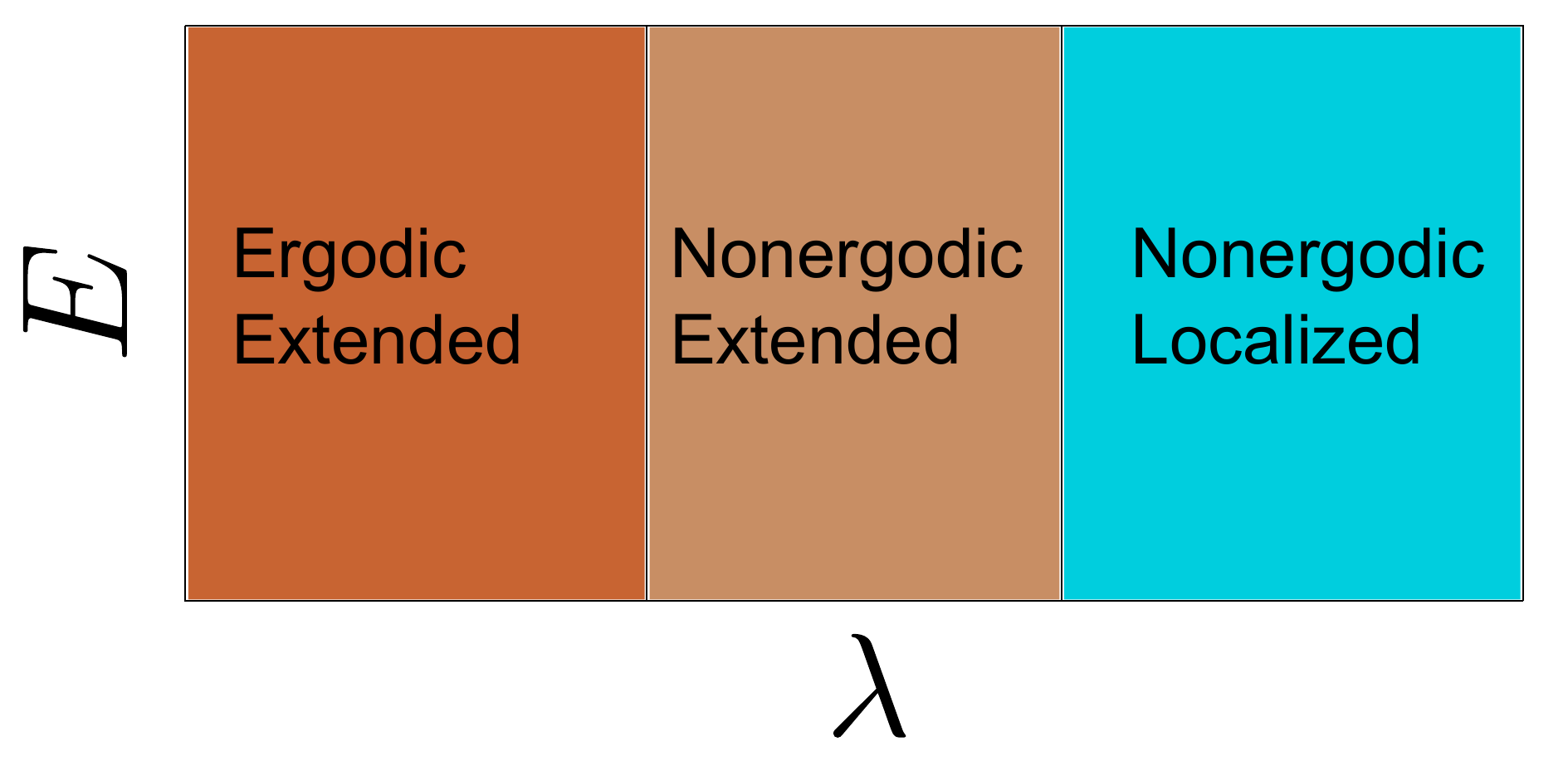}
 \caption{A scenario for many body localization transition. For non-interacting fermions in incommensurate lattice models, we have confirmed this phase transition scenario. Here $E$ is the many-body energy. The intermediate  ``nonergodic extended" phase shows up in the presence of a single-particle mobility edge. For interacting systems, i.e., the many body localization case,  if we assume an intermediate regime with a sub-extensive number of local integrals of motion, the scenario shown here also applies. 
}
\label{fig:LIMtransition} 
\end{figure}

\section{Perspectives from local integral of motion} 
\label{sec:LIM} 

We also consider localization and ergodicity breaking from the perspective of local integrals of motion (LIOM). For non-interacting fermions, in the presence of a single-particle mobility edge, we can take the occupation numbers of the localized orbitals as one set of local integral of motion. In this case, the number of LIOM is an extensive quantity. In the thermodynamic limit, the probability of occupying  the corresponding localized orbitals is finite for the whole energy range, as a consequence of statistics (Eq.~\eqref{eq:Pmalpha}). This immediately means that the non-ergodic states spread over the whole energy spectra, which is consistent with a more explicit construction in Ref.~\onlinecite{2015_Li_MBL_PRL}. 

If we start from a localized phase,  and  decrease the disorder strength, we would have three different regimes 
for a system with a single particle mobility edge 
(see Sec.~\ref{sec:GAALIM} for a concrete example). In a completely localized phase, we have the number of LIOM equal to the system size. In the intermediate regime, we have an extensive number of LIOM smaller than the system size, which makes the  spectra non-ergodic. In the completely extended phase,  the number of LIOM is zero. This transition scenario is shown in Fig.~\ref{fig:LIMtransition}. 
In a system without any mobility edge, the intermediate `nonergodic extended' phase does not appear.

For an interacting system, whether we have an intermediate regime with a sub-extensive (extensive but smaller than the system size) number of LIOM or not is  unknown.  But if we assume it exists, then the intuition and the scenario (Fig.~\ref{fig:LIMtransition}) from the non-interacting fermions should apply. In this intermediate regime, the interacting Hamiltonian would be block diagonal. The number of blocks is the dimension of the Hilbert space spanned by the set of LIOM, which is exponential in system size. Each block is presumably ergodic and obeys ETH. But the dimension of the block Hamiltonian is typically exponentially smaller than the total Hilbert space dimension, with the ratio between them approaching zero in the thermodynamic limit.  It follows that the eigenstates of the interacting system have all the properties of the non-interacting partially-extended states as described in Sec.~\ref{sec:NonErgodicPEP}. 
Our speculation is that turning on interactions does not immediately suppress the intermediate phase, and this is consistent with the finding in Ref.~\onlinecite{2015_Li_MBL_PRL,modak2015many}.

\subsection{Local integral of motion for the GAA model} 
\label{sec:GAALIM} 

Here, as a concrete example, 
we derive local integrals of motion for the GAA model and study its connection to mobility edge physics. 
Following Ref.~\onlinecite{2015_Modak_LIM_arXiv}, we construct LIOMs corresponding to the GAA model with the mobility edge. The models considered in the Ref.~\onlinecite{2015_Modak_LIM_arXiv} did not contain a mobility edge in the energy spectrum. It is  {\it a priori} not obvious if and how LIOMs 
are associated with 
a mobility edge.  Before constructing LIOMs, we consider the Hamiltonian in the diagonal form in the basis of eigenstate projectors $|\psi_i\rangle\langle\psi_i|	$,
\begin{align}
H=\sum^N_{i=1} \epsilon_i |\psi_i\rangle\langle\psi_i|	. 
\end{align}
Here, we have defined $\epsilon_i$ as the eigenvalue. If the system manifests a mobility edge, there will be a critical energy $\epsilon^*$ that separates the localized and extended states. In this basis,  projectors $|\psi_i\rangle\langle\psi_i|$ 
correspond to the LIOMs associated with the localized state. Notice that the projectors $|\psi_i\rangle\langle\psi_i|$ corresponding to the delocalized states 
are not local in nature even though they are integrals of motion.   
Such projectors can be constructed numerically using exact diagonalization methods. 

In the following we construct explicit local conserved charges $q(l=0$, \ldots, $N-1)$ for the GAA model following a recursive procedure~\cite{2015_Modak_LIM_arXiv} 
that are related to the projectors $|\psi_i\rangle\langle\psi_i|$ by a gauge transformation. Below, we check if local conserved charges are reliable indicators of localization for a model with a mobility edge. 

We start with the GAA Hamiltonian in Eq.~\eqref{eq:GAA} now rewritten as, 
\begin{align}
H &=\sum_i h_i n_i-y\sum_{ij}t c^{\dagger}_ic_{i+1}+h.c . 	
\label{model}
\end{align}
For the non-zero hopping parameter $t$ one can construct conserved charges ${q}(l)$ systematically in powers of $y$. The convergence of the power series in $y$ determines the existence of local conserved charges.  Consider the zeroth order conserved charge ${q}_0(l)$ corresponding to the zero hopping limit which is simply the onsite density $n_0$. When we introduce a nearest neighbor hopping term, then ${q}_0(l)$ can be expressed as an expansion in the hopping parameter $y$ and $t$, which does not truncate in the thermodynamic limit. Localization and delocalization are encoded in the convergence or divergence of this power series. Consider the following form for ${q}_0(l)$,
\begin{align}
{q}_0(l)=P_0(l)+y P_1(l)+y^2 P_2(l)+.....\\
P_0=n_0, \, \ \ P_m=\sum_{ij}\eta^m_{ij}(l)(c^{\dagger}_ic_j+c^{\dagger}_jc_i)	
\label{q0}
\end{align}
The operator $P_m$ takes a most general quadratic form. The coefficients $\eta^m_{ij}(l)$ are recursively evaluated 
enforcing
commutation of the charges with the Hamiltonian. This commutation condition imposes constraints on the coefficients in the form of a recursion relation. To detect the mobility edge, we need to compute all charges ${q}_0(l=0..N-1)$,
\begin{align}
[{q}_0(l), H]=[P_0(l), H]+&\sum^{\infty}_{m=0}y^{m+1}\\
\times &([P_m(l), H_1]+[P_{m+1}(l),H_0])\nonumber\\
\eta^{m}_{ij}=\frac{t}{h_j-h_i}(\eta^{m-1}_{i,j-1}+\eta^{m-1}_{i,j+1}&-\eta^{m-1}_{j,i-1}-\eta^{m-1}_{j,i+1}), \ \forall (i<j)
\end{align}
The above recursion in $\eta^m(l)$ can be computed order by order for all terms given some initial conditions. The recursion structure is symmetric for different charges except at the initial condition,
\begin{align}
 \eta^0_{ll}=1,\ \ \  \eta^m_{ij}(l)=\eta^m_{ji}(l), \ \ \eta^0_{ij}(l)=0 \,  (\forall i\ne0,\ j\ne 0).
\end{align}
This recursion generates growing number of hopping terms with increasing order of expansion $m$. The convergence of the power series is indicated by the typical term $\eta^m_{l,m+l}$.

\begin{figure*}[htp] 
\includegraphics[width=.8\textwidth]{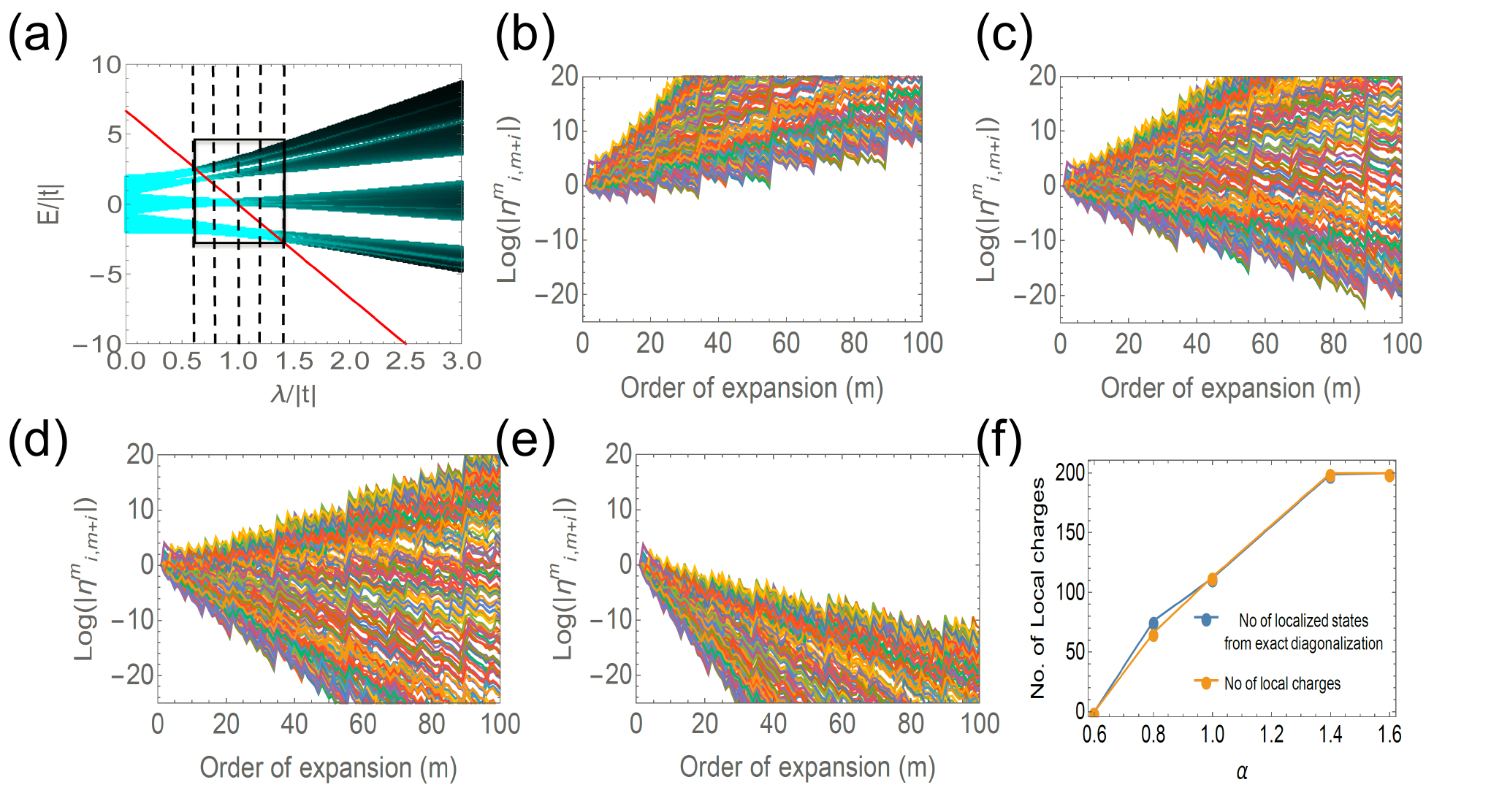}
\caption{Local integrals of motion in the GAA model. (a), the numerical energy spectrum $E/|t|$ as a function of $\lambda/|t|$ for $L=200$ sites tight binding model for $\alpha=0.3$. Pure cyan denotes IPR=0 and pure black denotes IPR=1. Red line is a plot of the analytically obtained mobility edge. Convergence of LIOMs: Plots of  $\log|\eta^m_{l,m+l}|$ ($l=0,...,199$) for different values of $\lambda$ ($\lambda/|t|=0.6,\, 0.8, \, 1.0, \, 1.4$, for (b), (c), (d) and (e)).
(f) shows comparison between the number of localized states obtained by exact diagonalization to the number of convergent local charges. 
}
\label{fig:liom1} 
\end{figure*}

We now test the convergence of local charges at different slices of the GAA spectrum with the mobility edge. 
Fig.~\ref{fig:liom1}(a) corresponds to the band spectrum of the GAA model plotted as a function of the potential strength $\lambda/|t|$ for $\alpha=0.3$. 
The red line corresponds to the analytical mobility edge $\alpha E=2\text{sgn}(\lambda)(|t|-|\lambda|)$. The boxed region corresponds to the region where the analytical mobility edge cuts the spectrum. The spectrum outside this boxed region does not manifest a mobility edge and all the energy states are either localized or delocalized. We plot the spectrum along with the IPR. For a localized eigenstate,  the IPR approaches the maximum possible value $\sim1$ (denoted by pure black). For an extended state, the IPR is of the order $1/L$, which is vanishingly small in the large system size limit (denoted by pure cyan). 

In Fig.~(\ref{fig:liom1}), we show the convergence of a typical term $\log|\eta^m_{l,m+l}|$ for all $l=0,..,199$ charges as a function of $m$ for a $N=200$ site system with periodic boundary conditions. The convergence of this term can be used to diagnose the convergence or divergence of a given charge. We carry out this calculation for four suggestive slices of $\lambda/|t|=0.6, 0.8, 1.0, 1.4$. 
Fig.~\ref{fig:liom1}(b) shows that the typical term $\log|\eta^m_{l,m+l}|$ diverge for all $l=0,..,199$ charges corresponding to $\lambda/|t|=0.6$. This is consistent with the numerical IPR result which indicates that all the states are extended. 
Fig.~\ref{fig:liom1}(c) shows the interesting case of the mobility edge in the spectrum  when $\lambda/|t|=0.8$. 
For this case, some typical terms ($\log|\eta^m_{l,m+l}|$) converge indicating the existence of local charges.
This is indicative of the existence of some localized states in the spectrum  as shown in the numerical IPR plot. Notice that there are more extended states than localized states in agreement with the majority of typical terms diverging. There are also some marginal typical terms which may converge or diverge decisively in the thermodynamic limit.  
Fig.~\ref{fig:liom1}(d) corresponds to the case of mobility edge for $\lambda/|t|=1$. Notice that for this case the majority states are localized and this is consistent with majority typical terms converging. The two mobility edge cases of $\lambda/|t|=0.8, 1$ clearly indicate that the number of local charges are directly proportional to the number of localized states. To obtain a quantitative agreement with the number of local charges and number of localized states, we need to go to larger and larger system sizes and higher order $m$ to determine the fate of the marginal charges. However, there is an excellent qualitative agreement. Finally, we consider the case of $\lambda/|t|=1.4$ where all the states are localized. 
Fig.~\ref{fig:liom1}(f) clearly shows that all typical terms converge indicating that there exists a full set of local charges (LIOMs) associated with a fully localized spectrum.
We also demonstrate excellent agreement between the number of convergent local charges compared to  the number of localized states obtained by exact diagonalization (see Fig.~\ref{fig:liom1}(f)).

\section{Conclusion} 
\label{sec:summary} 

To summarize, we have investigated many-body effects of single-particle mobility edges 
on the full many-body spectra of noninteracting
incommensurate and disordered fermionic models 
in order to shed light on the issue of many-body-localization properties of generic interacting systems where the corresponding single-particle models (without interactions) have mobility edges separating extended and localized states. 
By comparing entanglement and thermal entropy as well as  calculating the non-thermal fluctuations of subsystem particle number, we distinguish between localization and quantum nonergodicity transitions. 
In particular, we show that models with single-particle mobility edges have two generic transitions (Fig.~\ref{fig:LIMtransition}) in their many-body energy spectra: one from the purely extended thermal phase with volume law entanglement entropy (which equals the thermal entropy) and thermal fluctuations to a partially extended nonthermal phase with subthermal volume law entanglement entropy (which is less than the thermal entropy) and extensive (i.e. nonthermal) particle number fluctuations, and the second one from the partially extended phase to the purely localized (with area law entanglement entropy) nonthermal phase.
We establish a nonergodic extended phase as a generic intermediate step interpolating thermal and localized phases of non-interacting fermions.  
We expect
that  the intermediate nonthermal partially extended phase should survive the situation with finite interaction, thus explaining the numerical findings of Ref.~\onlinecite{2015_Li_MBL_PRL} 
involving interacting fermions.  Thus, the presence of an intermediate exotic ``nonthermal metallic phase" between purely thermal and purely localized phases may very well be a generic feature of the physics of many body localization of interacting fermions.
For many-body fermionic states in non-interacting 1d Aubry-Andr\'e and 3d Anderson models, we have shown that the entanglement entropy density and the  normalized particle-number fluctuation  have discontinuous jumps (in the thermodynamic limit)  at the localization transition. We expect the  critical behaviors of particle-number fluctuations as established in this work to provide important guidelines for experiments to probe the localization and thermal-to-nonergodic transitions.
In addition to these results for the many-body localization properties of models with single-particle mobility edges, we also studied (section~\ref{sec:AAsingleparticle}) in depth the critical properties of the incommensurate potential driven localization transition in the generalized Aubry-Andr\'e  model (Eq.~\eqref{eq:GAA}) which manifests a mobility edge in one dimension.  We show that the universality class of the GAA model is the same as that of the AA model since both have the same correlation length exponent $\nu =1$.  We also provide detailed results for the level statistics, inverse participation ratio, and most importantly, the local integrals of motion for the incommensurate localization models, and establish that they are fundamentally different from Anderson localization models.

In conclusion, we comment on some features of the intermediate nonthermal extended MBL phase which appears to be generically present whenever there is a single-particle mobility edge.  In the appendix, we provide arguments for the generic existence of the intermediate `partially extended' phase in the noninteracting many-body spectrum, showing that in addition to many-body Slater determinant states consisting of purely extended and purely localized single-particle orbitals, it is unavoidable that the existence of the mobility edge in the single-particle spectrum would necessarily allow for many-body states with mixed localized and extended orbitals (leading to the partially extended intermediate phase).  This phase is thus unavoidable in the noninteracting fermionic many-body system if the corresponding single-particle model has a mobility edge.  We emphasize, however, that our arguments also show that such an intermediate many-body phase cannot arise (at least in the noninteracting problem) if the corresponding single-particle model does not have a mobility edge since all many-body product wavefunctions must then consist of either all extended or all localized single-particle states.  (We also note as an aside that our arguments apply only to fermions, not to bosons.)  Thus, the existence of the partially-extended many-body nonthermal phase with extensive entanglement entropy (which is not equal to the thermal entropy) is now an established fact for noninteracting fermionic many-body sates in models with single-particle mobility edges.  Numerical work shows~\cite{2015_Li_MBL_PRL} that this intermediate phase also exists in the presence of interaction, and indeed, we see no obvious reason for this phase to disappear for weak interactions. Macroscopic thermodynamic and transport properties of this partially-extended nonergodic metal phase are unknown at this stage except that this many-body phase has extensive subthermal entanglement and extensive (nonthermal) particle number fluctuations.  It may be worthwhile to mention in this context that in purely classical systems, a metal is characterized by diffusive transport behavior, but many instances are known for models with sub-diffusive transport where the root mean square distance the particle traverses grows in time as a power law smaller than $1/2$.   Thus, dynamic phases with behaviors in between diffusive metals and pure insulators are not uncommon even in classical systems.  
{Finally, we mention that, 
after Refs.~\onlinecite{2015_Li_MBL_PRL,modak2015many}, there have been recent efforts~\cite{2015_Rahul_PRB,2016_Bauer_arXiv} } 
to understand the role of single-particle mobility edges on many-body localization properties by dividing the interacting many-body system into a bath and a system which are comparable in size, and then arguing that the back action of the system on the bath could localize the bath so that it might be possible for true MBL to arise in systems with single-particle mobility edges.  The intermediate partially extended phase then might arise from some delicate bath-system interplay in such models, which however would be difficult to discern in finite size simulations.  Although such approaches are promising, in principle, there are aspects of the models we study which are not captured in such naive bath-system dichotomy.   In particular, the mobility edge divides the single-particle spectrum in a precise manner between extended and localized states in energy, and this is difficult to capture in an independent bath-system model,  where the separation is in real space.  Also, the precise role of fermions in the problem (see Appendix) remains obscure in such a bath-system division where bosons and fermions should be equivalent. Our current study invoking noninteracting many-body properties clearly establishes the generic existence of the intermediate non-ergodic metallic phase as well as the two-step MBL transitions in models with single-particle mobility edges, and thus, we believe that the corresponding interacting system should also manifest the intermediate phase as has been found numerically~\cite{2015_Li_MBL_PRL}.  More work is obviously necessary to decisively establish the existence of the intermediate MBL phase in the presence of interactions since small system sizes used in the exact diagonalization studies of interacting systems~\cite{2015_Li_MBL_PRL} suffer from serious finite size limitations which our noninteracting studies using very large system sizes do not suffer from.

\section*{Acknowledgement} 
This work is supported by JQI-NSF-PFC, ARO-Atomtronics-MURI, LPS-MPO-CMTC, and Microsoft Q. XL would like to thank Y.-Z. You for helpful discussions. The authors acknowledge the University of Maryland supercomputing resources~\cite{umdhpcc} made available for conducting the research reported in this paper.

\appendix 

\section{Many-body states of non-interacting fermions in the presence of a single-particle mobility edge} 
\label{sec:apppartext}

\begin{figure}[htp] 
\includegraphics[angle=0,width=\linewidth]{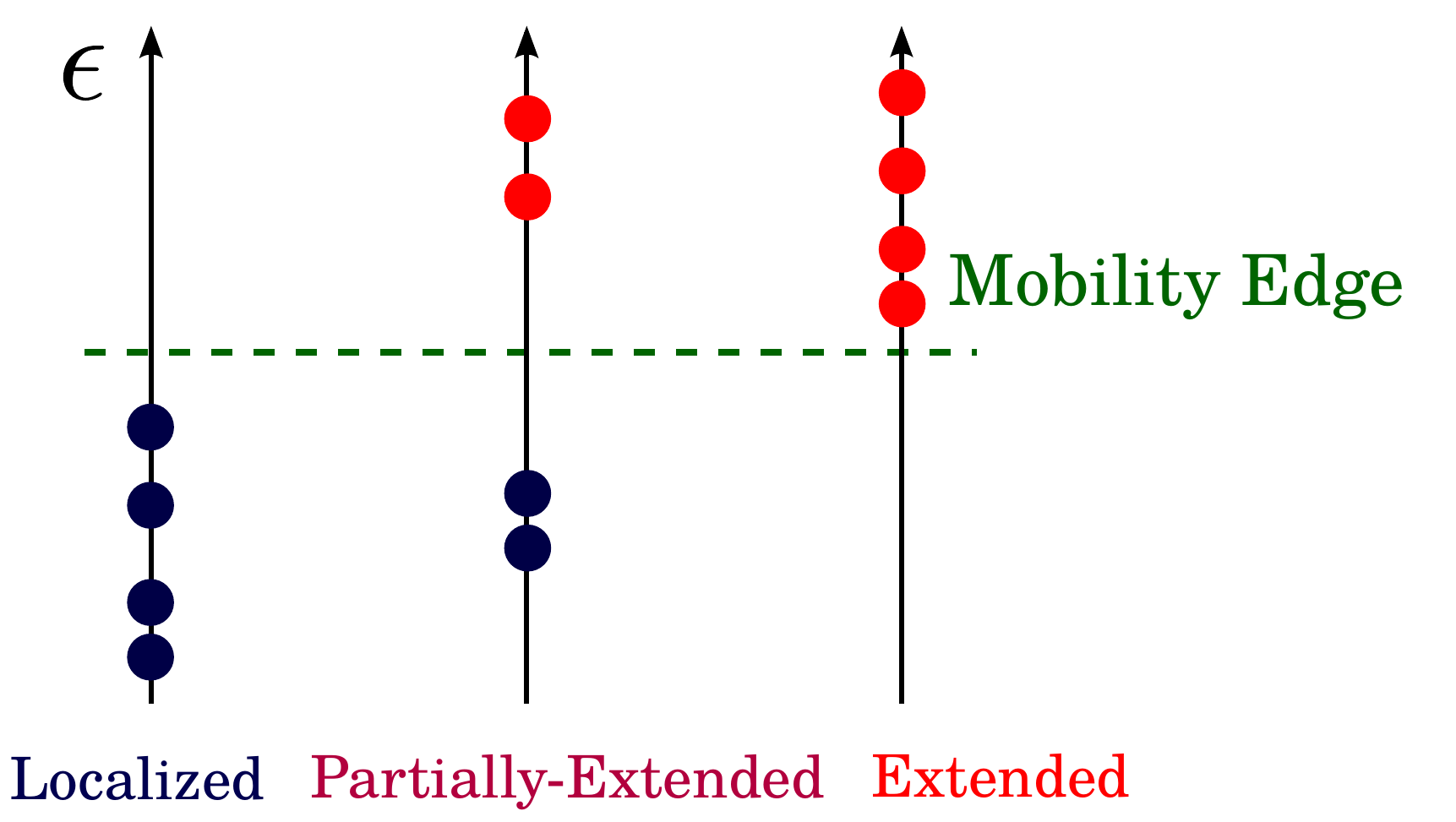}
 \caption{Pictorial illustration of three different possibilities to construct many-body states of non-interacting fermions in the presence of a single-particle mobility edge. Here $\epsilon$ is the single-particle energy axis. The `dashed' line marks the single-particle mobility edge. In this plot, we assume the single-particle states below (above) the mobility edge are localized (delocalized). With all particles put below the mobility edge, the many-body state is localized (shown on the left) with area-law entanglement entropy. With all particles put above the mobility edge, the many-body state is extended (shown on the right) with volume-law entanglement entropy, which is equal to the thermal entropy. Another scenario is to put some particles below the mobility edge and some above it, which leads to a partially extended state (shown in the middle). This state has volume-law entanglement entropy, yet smaller than the thermal entropy.}
\label{fig:pextpictorial} 
\end{figure}

 Without interactions, the many-body eigenstate of $N$ fermions is a product state of $N$ single-particle orbitals, 
 \be 
 |\Psi\rangle_{\rm free} = |m_1, m_2, \ldots, m_N\rangle = \psi_{m_1} ^\dag \psi_{m_2} ^\dag \ldots \psi_{m_N} ^\dag |{\rm vac} \rangle, 
 \ee 
 with $\psi_m^\dag $ a creation operator for a single-particle eigenstate. 
 Consider a model with single-particle energies $\epsilon_1 < \epsilon_2< \ldots < \epsilon_L$ ($L$ is the system size) having a single-particle mobility edge $\epsilon_{m^*}$, such that the states with $\epsilon_{m \leq m^* } $ ($\epsilon_{m>m^*}$) are localized (delocalized). In the presence of this single-particle mobility edge, there are three possibilities to construct a many-body state, namely localized, extended, and partially extended states  (see Fig.~\ref{fig:pextpictorial}). A many-body partially-extended state (Fig.~\ref{fig:pextpictorial}) with the lowest energy we can construct is then given by 
 $$ 
 |\Psi \rangle _{\rm partial-ext} ^{\rm low}  = |1, 2, \ldots, N-1, m^* +1  \rangle, 
 $$  
 which has a total energy 
 \be 
 E_A = \epsilon_{m^* +1} + \sum_{m = 1} ^{N-1} \epsilon_m . 
 \ee 
 A partially-extended state with the highest energy is 
 $$
 |\Psi \rangle_{\rm partial-ext} ^ {\rm high} = | m^*, L -(N-2), L-(N-3), \ldots, L\rangle, 
 $$  
with  a total energy 
 \be 
 E_B  =\epsilon_{m^*} + \sum_{m = L-(N-2)} ^{L} \epsilon_m. 
 \ee 
 For a lattice model, the energy spectrum is bounded, say between $(E_{\rm ground}, E_{\rm upper})$. 
 With non-interacting fermions, we have $E_{\rm ground}  = \sum_{m=1} ^N \epsilon_m$, and  
 $E_{\rm upper} = \sum_{m = L - (N-1)} ^{L} \epsilon_m$. 
In the energy windows $E\in (E_{\rm ground}, E_A)$, and $E \in (E_B, E_{\rm upper})$, we only have localized and extended many-body states, respectively,  whereas in the energy window   $E\in (E_A, E_B)$, we have partially-extended states coexisting with localized and extended states. 
 In the thermodynamic limit, the energy ranges $E_A - E_{\rm ground}$, and $E_{\rm upper} - E_B$ are  intensive, and the energy range $E_B - E_A$ is an extensive quantity. Thus the coexisting energy window completely dominates the energy spectrum in the thermodynamic limit.

\begin{figure*}[htp] 
\includegraphics[angle=0,width=\linewidth]{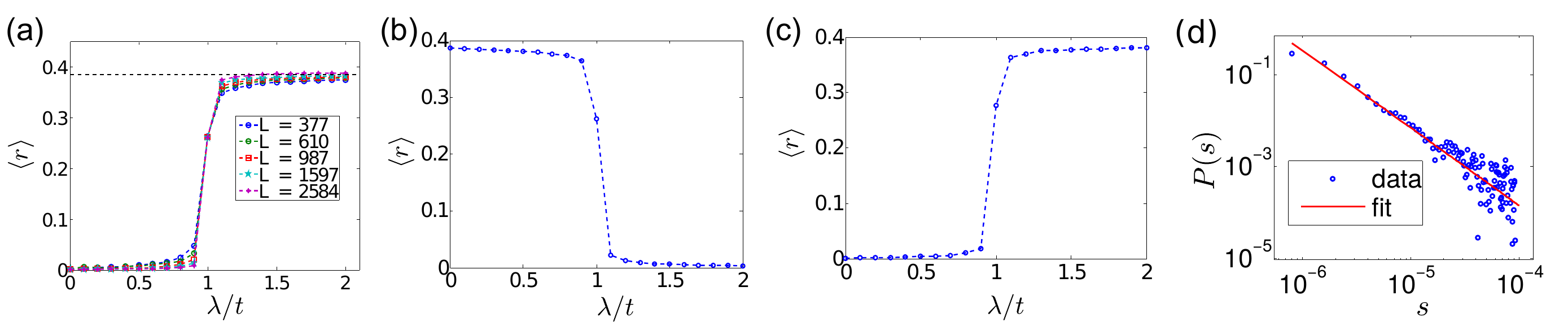}
 \caption{Level statistics of the AA model. We average over all single particle eigenstates here. In (a), we average over $\phi$. The black dashed line in (a) marks the position of $2{\rm ln}2 -1$, an expected value from Poisson statistics. As approaching the thermodynamic limit, $\langle r\rangle$ approaches to zero in the delocalized phase ($\lambda/t <1$) and to $2{\rm ln}2-1$ in the localized phase ($\lambda/t>1$). 
 In (b), we average over  $\varphi$ (flux), and the system size is $L = 987$. In both (a) and (b) $Q$ is set from the Fibonacci sequence as described in the main text. In (c) $Q$ is set from the Pell sequence and we average over $\phi$ as in (a).   The results imply that $\langle r\rangle $ value does not depend on the choice of  irrational numbers for the wavenumber $Q$. 
 (d) shows the probability distribution (with double logarithmic scale) of the level spacing of the AA model at the self-dual point. In (d), we use $\lambda/t = 1$, $L = 4181$, $Q/2\pi = 2584/4181$, the shift $\phi$ is averaged over. The distribution $P(s)$ fits well to a power-law function  $\sim s^{-1.68}$. 
 }
\label{fig:rvalueAAH} 
\end{figure*}

\section{Single-particle properties of incommensurate lattice models} 
\label{sec:singleparticle}


\subsection{Duality of the Aubry-Andr\'e model and the level statistics} 
In this section, we review the duality of the AA model~\cite{AA, azbel, harper55}, 
i.e. the model obtained by putting $\alpha=0$ in $H_0$ of Eq.~\eqref{eq:GAA}, 
and discuss its consequences on level statistics. 
The Hamiltonian of the AA model with twisted boundary conditions is $H = H_t + H_{\lambda}$, 
\bea
&& H_t = \sum_{j =0} ^{L-1} \left[  -t e^{i \varphi } c_j ^\dag c_{j+1} + h.c. \right]  \\ 
&& H_{\lambda} = -2\lambda \sum_{j=0} ^{L-1} \cos (Q j + \phi) c_j ^\dag c_j. 
\eea  
The Hamiltonian is a function of the parameters $t$, $\varphi$, $\lambda$, and $\phi$---$H[t, \varphi; \lambda, \phi]$. 

The duality of the AA model can be seen from the Fourier transform 
\be 
c_j = \frac{1}{\sqrt{L}} \sum_{k = \frac{2n\pi}{L}} c(k) e^{i k j}. \nn 
\ee 
In the momentum space, we have 
\bea
H_t = -2t \sum_{k} \cos (k + \varphi) c^\dag (k) c(k) \nn  \\ 
H_\lambda = -2 \lambda \sum_{k, k'} V(k,k') c^\dag (k) c(k'),    \nn 
\eea 
with 
\be 
V(k,k') = \frac{1}{2L} \sum_j \left\{ e^{ i \left[ (k'-k+Q)j + \phi \right]} + e^{ i \left[ (k'-k-Q)j-\phi\right] } \right\}. \nn  
\ee 
In a finite size system, to have the duality, we need to take 
\be 
Q = \frac{2\pi M }{L},
\label{eq:Q} 
\ee 
where  $M$ is an integer.  The other requirement is $M$ and $L$ are co-primes.

Given Eq.~\eqref{eq:Q}, we have 
$$  
V(k,k') = \frac{1}{2} \left\{ 
				e^{ i\phi} \delta_{k'-k +Q, 0} + e^{- i\phi} \delta_{k'-k-Q,0} \right\}. 
$$  
Then it follows that 
\be 
H_{\lambda} = \sum_k \left[ -\lambda e^{i\phi} c^\dag (k) c(k-Q) -\lambda e^{-i\phi} c^\dag (k) c(k+Q) \right]. \nn 
\ee 
We can relabel the momentum by $ k = m Q$ mod $2\pi$, with $m = 0, 1, \ldots (L-1)$. This relabeling relies on that $M$ and $L$ are co-primes (otherwise, this relabeling cannot be done).  
With this relabeling, $c(k) \to \tilde{c}(m)$, we have 
\bea 
&& H_{\lambda} = \sum_{m} \left[ -\lambda e^{i\phi} \tilde{c}^\dag (m) \tilde{c} ( m-1)  + h.c.\right]  \nn \\ 
&& H_{t}  = -2 t \sum_m \cos (Qm + \varphi) \tilde{c}^\dag (m) \tilde{c} (m). \nn 
\eea 
We thus conclude that $H [t, \varphi; \lambda, \phi]$ is equal to (up to a unitary transformation) $H^*[ \lambda, \phi; t, \varphi]$, i.e., 
\be 
H [t, \varphi; \lambda, \phi] = U^\dag  H^*[ \lambda, \phi; t, \varphi] U .
\label{eq:AAHduality} 
\ee  

Under the duality transformation, the delocalized states map to localized states, but the energy spectrum stays the same. Thus the level statistics for the localized phase is the same as the delocalized phase in the AA model. One can make the level statistics appear differently for the two phases by selectively averaging over $\phi$ or $\varphi$.

Here we give two examples of setting $Q$ and $L$ in a finite size system. One is based on Fibonacci numbers defined by 
\be 
F_0 = 1, F_1 = 1, F_n = F_{n-1} + F_{n-2}. 
\ee 
We can set $L = F_n$, $Q/2\pi = F_{n-1}/F_{n}$ ($F_{n-1}$ and $F_n$ are co-prime). As $n\to\infty$, $Q/2\pi$ approaches the inverse golden ratio $2/(1+\sqrt{5})$. 
Alternatively, one can also use Pell numbers defined by 
\be 
P_0 = 0, P_1 = 1, P_n = 2 P_{n-1} + P_{n-2}. 
\ee 
In this case, as $n\to\infty$, $Q/2\pi = P_{n-1} /P_n$ approaches the inverse silver ratio $\sqrt{2} -1$. 

Alternatively for a large system, it is not necessary to maintain the precise duality in numerical calculations to determine the localization transition, and we can directly set the wavenumber $Q$ to be an irrational number. 

To characterize the level statistics, we calculate the adjacent gap ratio $r$, defined to be 
\be 
r_n = \frac{ {\rm min} (s_n, s_{n-1})}{ {\rm max} (s_n, s_{n-1}) }, 
\ee  
with $s_n$ the level spacing between the $n$th and $(n-1)$th eigenstates. The average of $r_n$ is 
introduced as $\langle r \rangle = \frac{1}{L} \sum_n r_n$. 
In Fig.~\ref{fig:rvalueAAH}, we show the level statistics across the localization transition. The distribution of level spacings at the self-dual critical point shows a power-law decay.

\begin{figure}[h]
\centering
\begin{minipage}{.25\textwidth}
  \centering
  \includegraphics[width=0.7\linewidth,angle=-90]{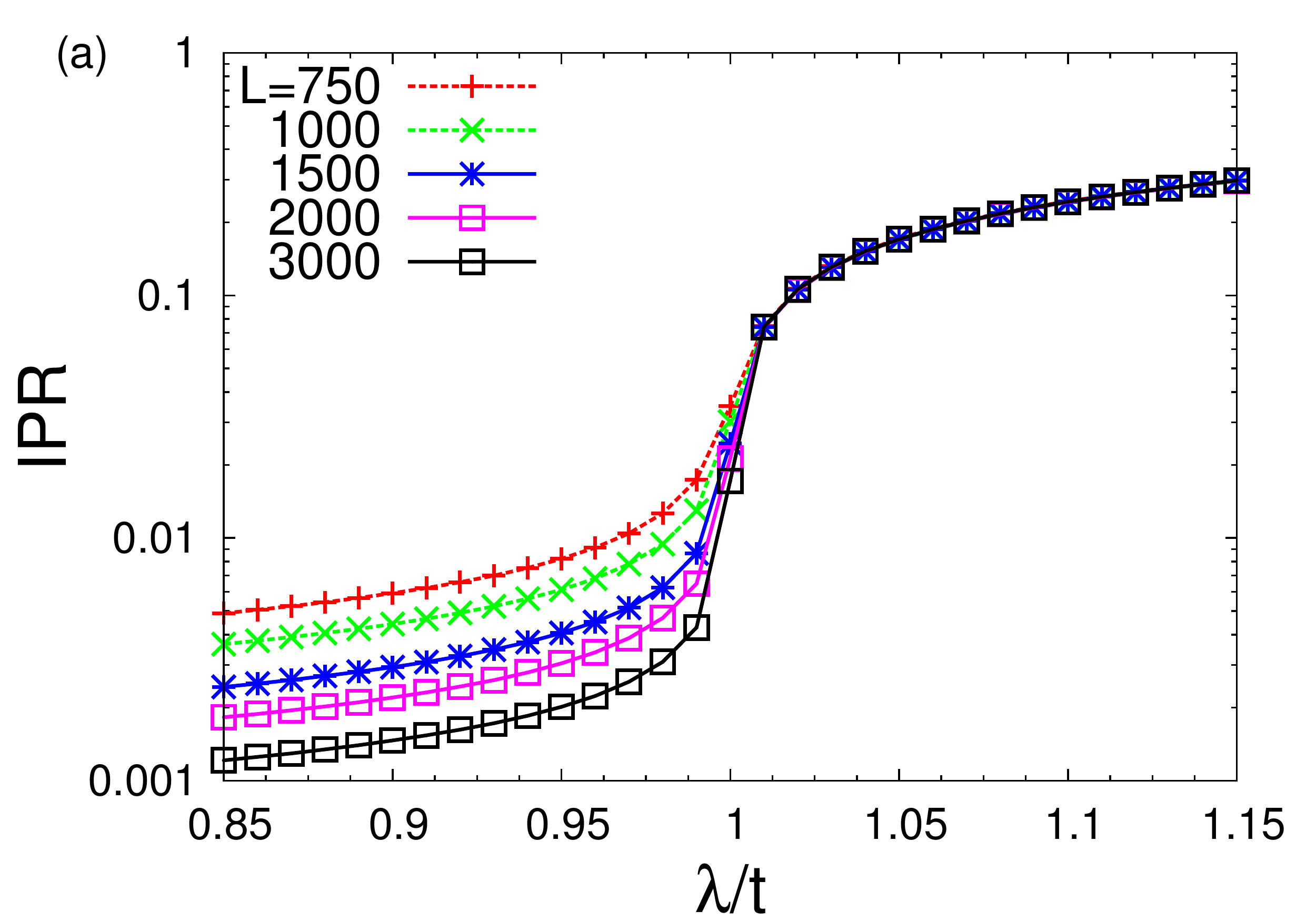}
\end{minipage}%
\begin{minipage}{.25\textwidth}
  \centering
  \includegraphics[width=0.7\linewidth,angle=-90]{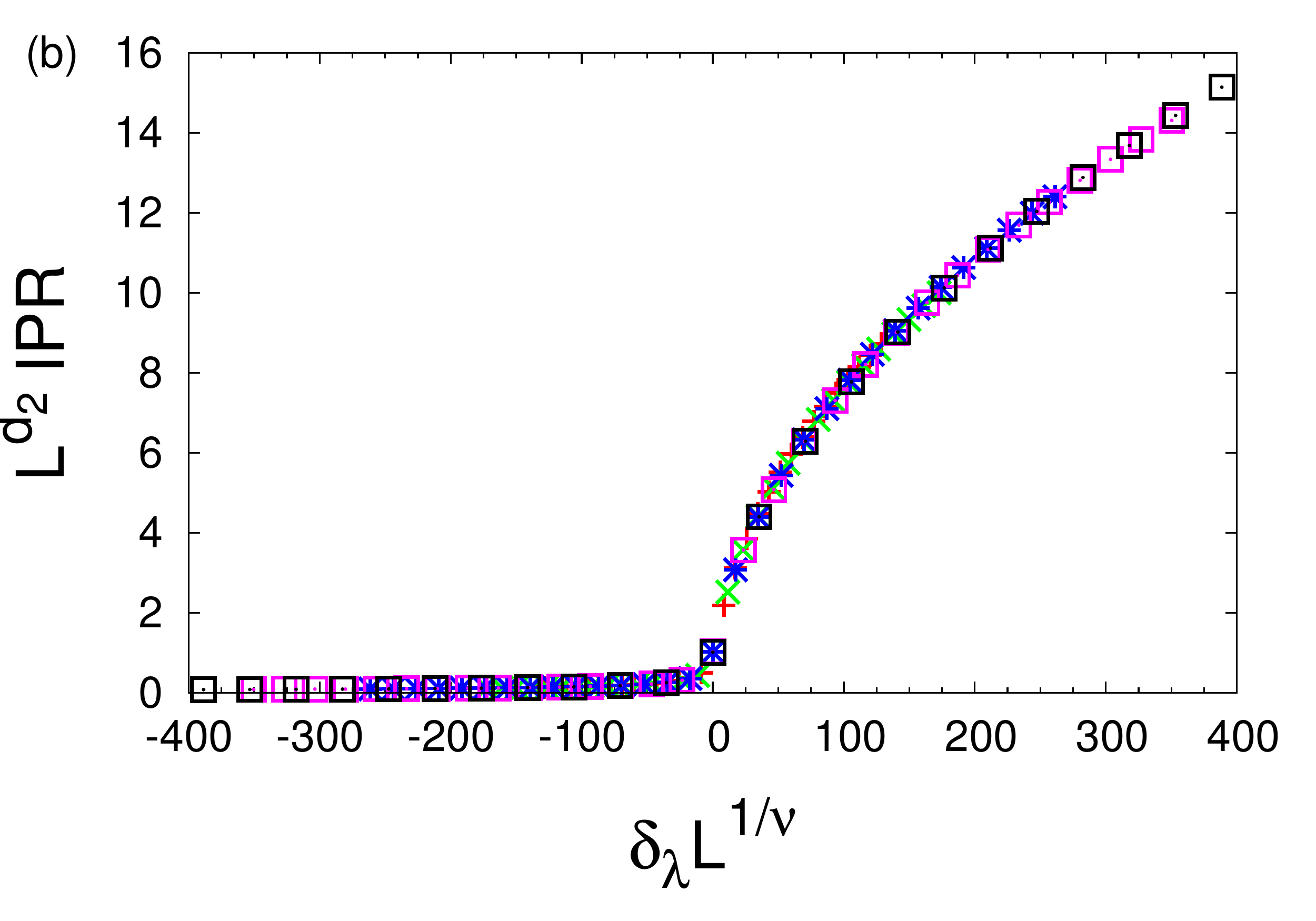}
\end{minipage}
\begin{minipage}{.25\textwidth}
  \centering
  \includegraphics[width=0.7\linewidth,angle=-90]{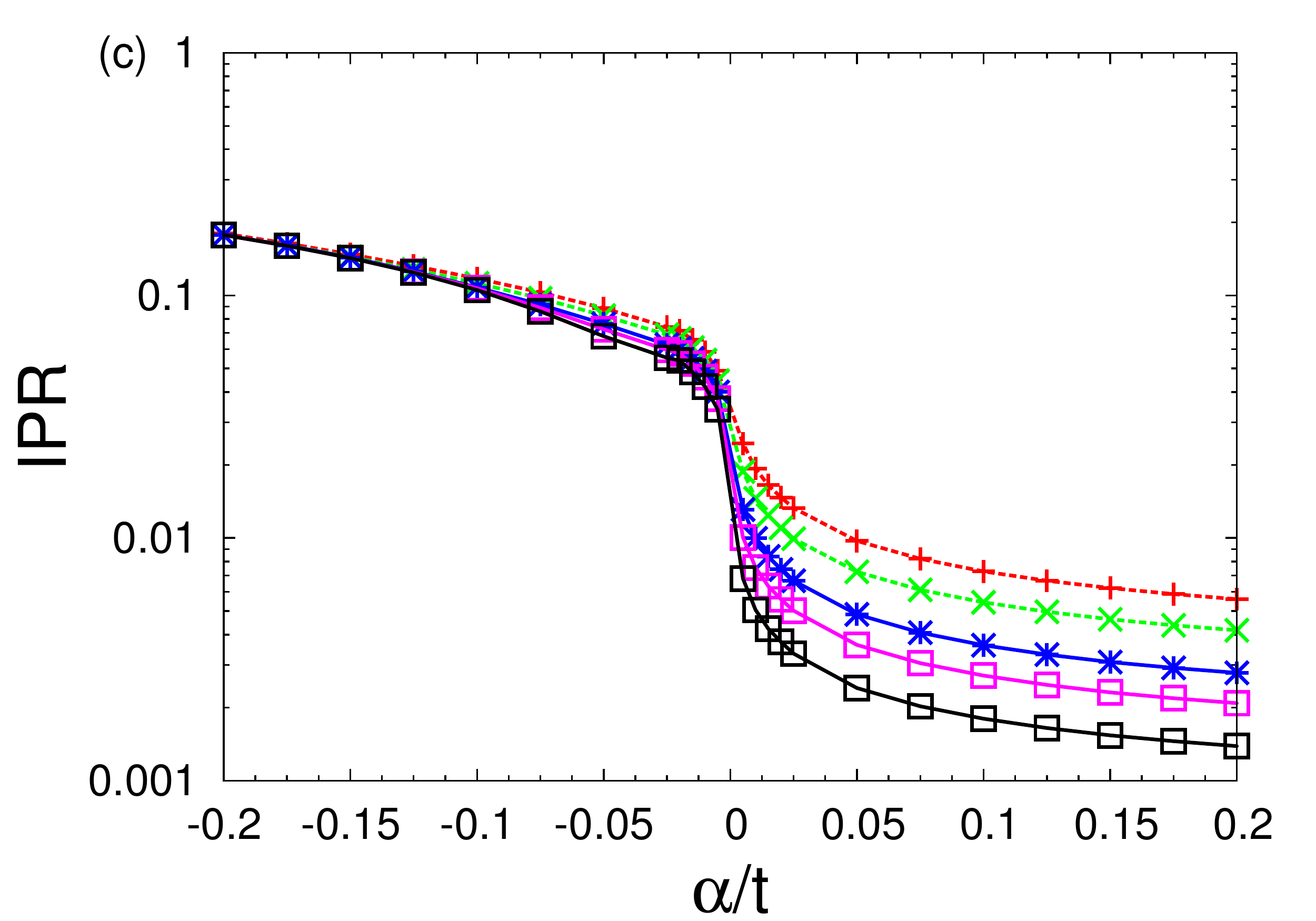}
\end{minipage}%
\begin{minipage}{.25\textwidth}
  \centering
  \includegraphics[width=0.7\linewidth,angle=-90]{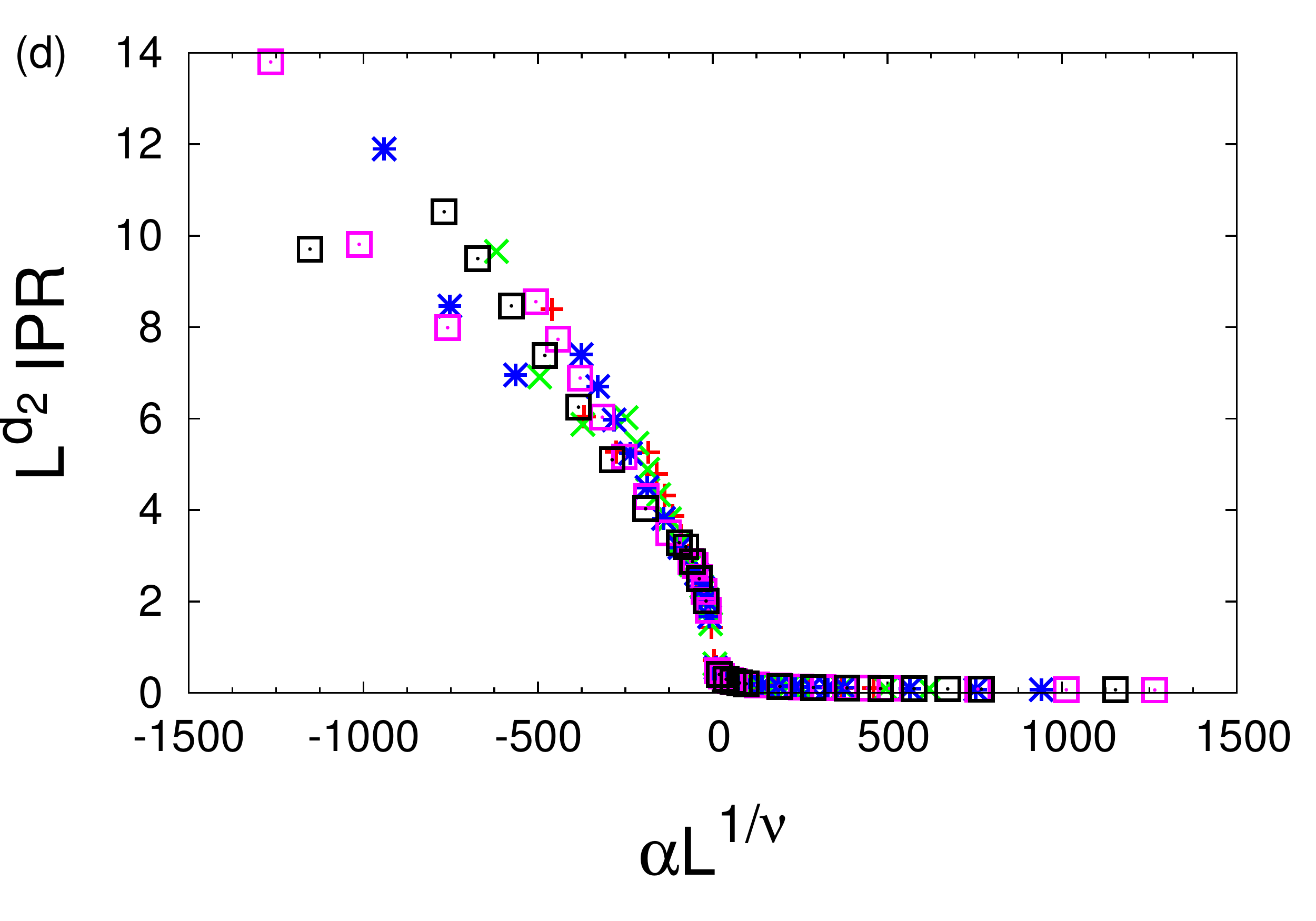}
\end{minipage}
\caption{Inverse participation ration near the localization transition for $\alpha=0$ in (a) and (b), and for $\lambda=\lambda_c$ in (c) and (d). The labels for each system size are shared across each figure. We find the IPR goes like $1/L$ in the delocalized phase and saturates to an $L$ independent constant in the localized phase.  Applying scaling collapse in the vicinity of the critical point yields the critical exponents given in Table~\ref{tab:exponents}. 
}
\label{fig:AA_IPR}
\end{figure}

\begin{figure}[h]
\centering
 \includegraphics[width=\linewidth]{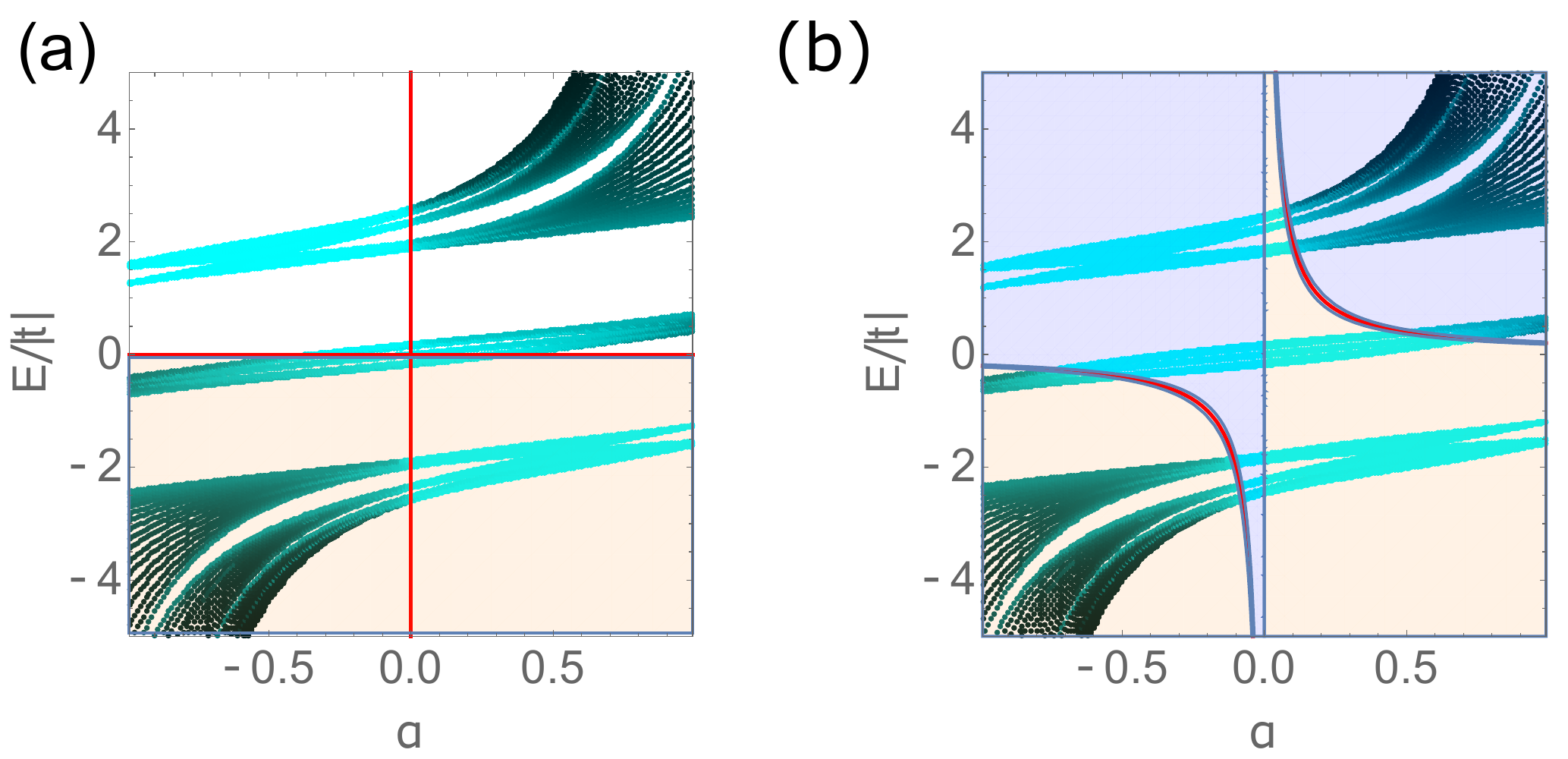}
\caption{Energy ranges averaged over when computing the IPR. The numerical single particle energy spectra is depicted with its corresponding IPR value that is not averaged over energy, black denotes an IPR=0, cyan denotes an IPR=1, and each red line marks the single particle mobility edge $\epsilon_{ME}$.   (a) $\lambda=t$, we average over all energies $\epsilon <0$ (the light tan region) giving rise to a localization transition at $\alpha=0$ shown in Fig.~\ref{fig:AA_IPR} (c) and (d). (b) $\lambda = 0.9t$ for energies $\epsilon < \epsilon_{ME}$ we average over the light tan region in energy whereas for  $\epsilon > \epsilon_{ME}$ we average over the light blue region of energy, resulting in the average IPR in Fig.~\ref{fig:GAA_IPR}. The localization transition in each case occurs where the single particle spectrum intersects the mobility edge which yields $\alpha_c^- = −0.07286$ and $\alpha_c^+ = 0.07631$.
}
\label{fig:GAAavescheme}
\end{figure}

\begin{figure}[htp]
\centering
\begin{minipage}{.25\textwidth}
  \centering
  \includegraphics[width=0.7\linewidth,angle=-90]{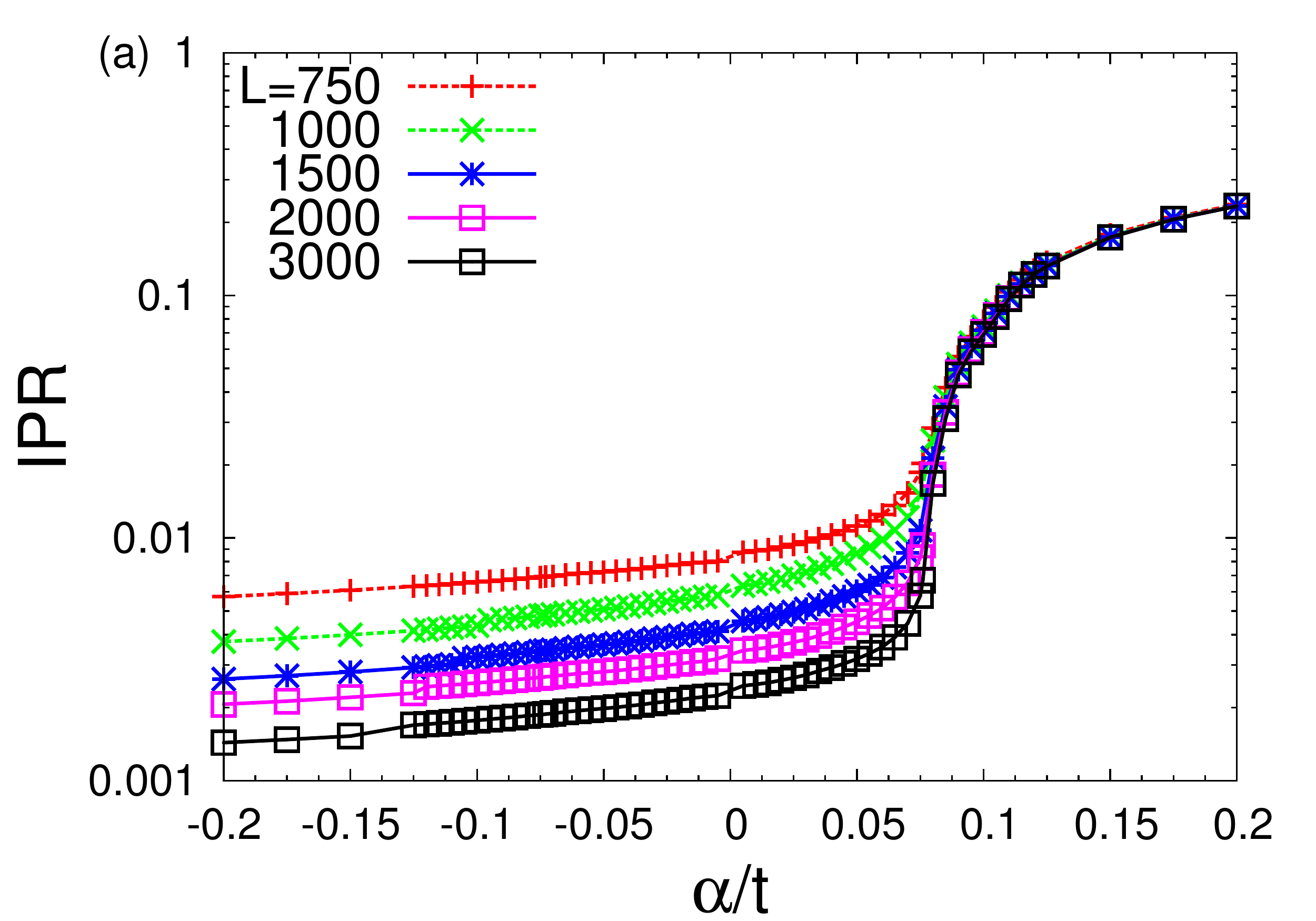}
\end{minipage}%
\begin{minipage}{.25\textwidth}
  \centering
  \includegraphics[width=0.7\linewidth,angle=-90]{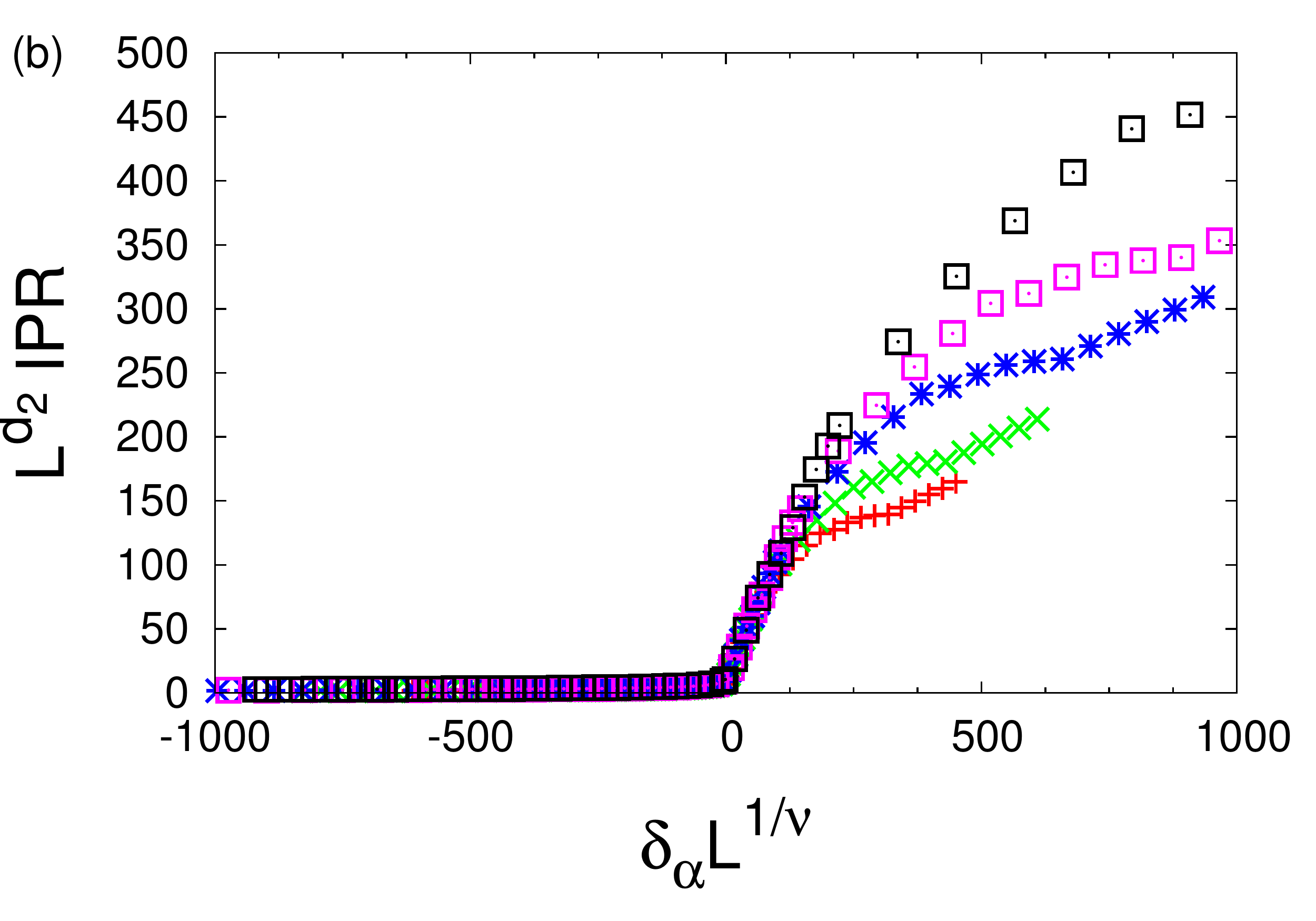}
\end{minipage}
\begin{minipage}{.25\textwidth}
  \centering
  \includegraphics[width=0.7\linewidth,angle=-90]{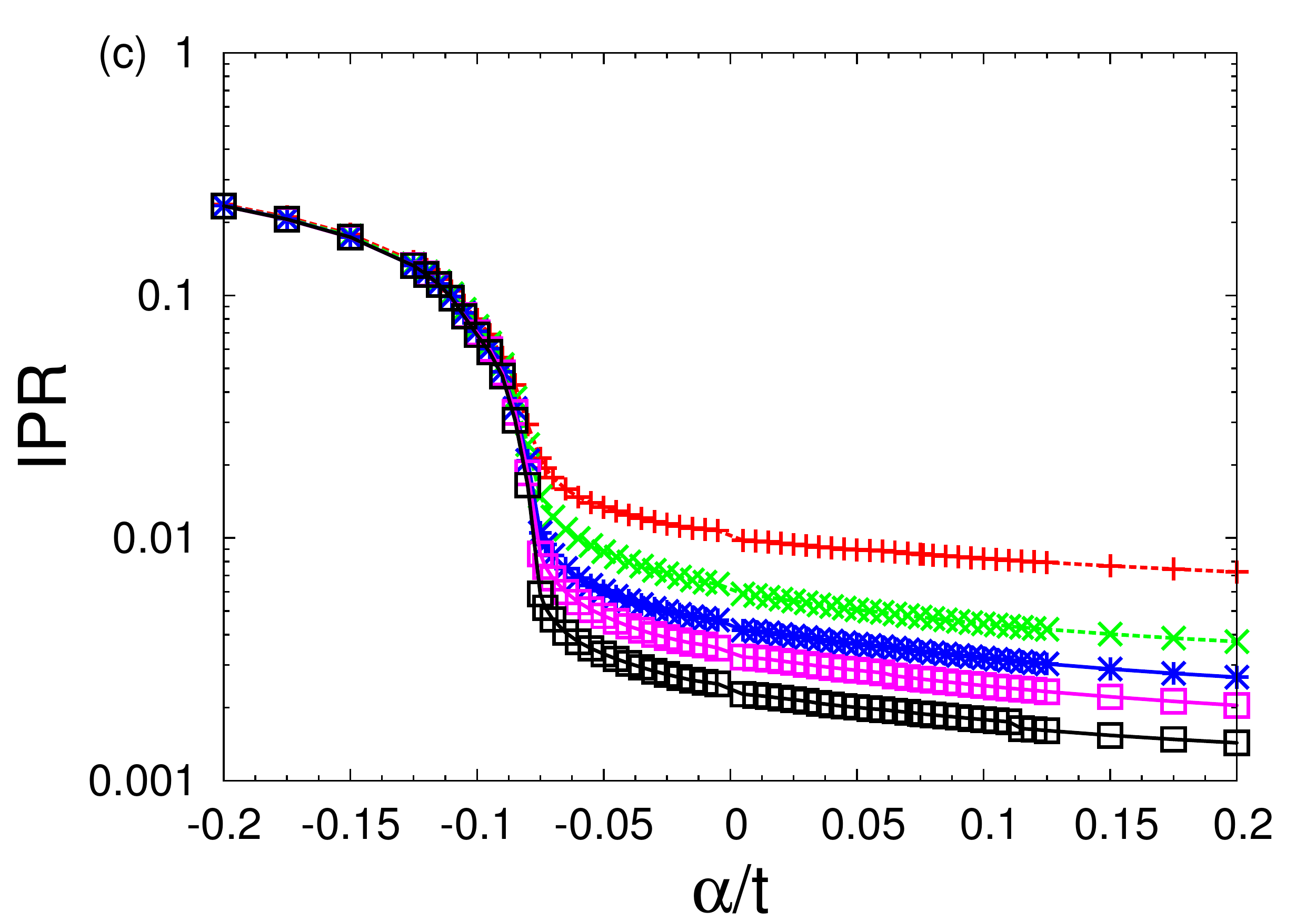}
\end{minipage}%
\begin{minipage}{.25\textwidth}
  \centering
  \includegraphics[width=0.7\linewidth,angle=-90]{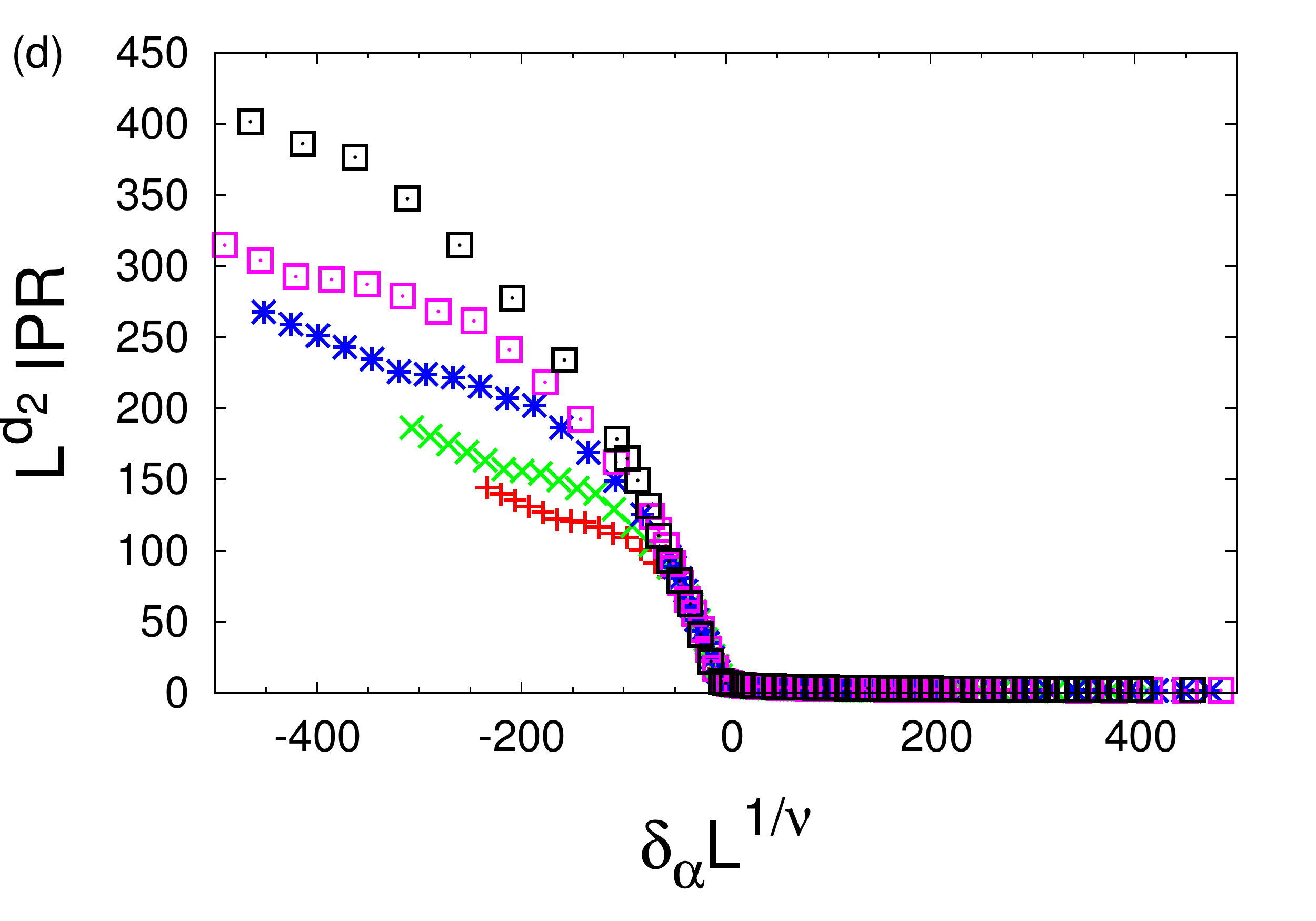}
\end{minipage}
\caption{Inverse participation ration near the localization transition for $\lambda=0.9t$, and averaging over energies $\epsilon>\epsilon_{ME}$ in (a) and (b), and averaging over energies $\epsilon<\epsilon_{ME}$ in (c) and (d). The labels for each system size are shared across each figure. The results for the critical exponents are given in Table~\ref{tab:exponents}. 
}
\label{fig:GAA_IPR}
\end{figure}

\subsection{Critical properties of the single-particle incommensurate lattice models}
\label{sec:AAsingleparticle}
In this section we study the critical properties of the generalized Aubry-Andr\'e model with a focus on determining the localization length exponent $\nu$. This exponent describes the divergence of the localization length on approach to the transition as
$
\xi \sim \delta_{\lambda}^{-\nu},
$
where we have defined the distance to the critical point $\delta_{\lambda} \equiv |\lambda - \lambda_c|/\lambda_c$. 
We consider the inverse participation ratio (IPR)
\begin{equation}
{\rm IPR} (\epsilon) =\sum_x |\psi_\epsilon (x)|^4 ,
\end{equation}
for normalized wave functions $\psi_\epsilon (x)$ at an eigenenergy $\epsilon$. 
A related quantity is normalized participation ratio 
\be 
\eta = \frac{1}{ L^d \times  {\rm IPR}}. 
\ee 
 The IPR acts like an order parameter for the localization transition, where it is an $L$ independent constant in the localized phase and vanishes like $1/L$ in the delocalized phase.  Near the localization transition (via tuning the potential $\lambda$ or $\alpha$ in Eq.~\eqref{eq:GAA}) the IPR satisfies the finite size scaling form 
\begin{equation}
 {\rm IPR} (\epsilon) \sim \frac{1}{L^{d_2(\epsilon)}} f(\delta L^{1/\nu}), 
\end{equation}
where we have made it explicit that the correlation length exponent is independent of the energy $\epsilon$, however the exponent $d_2$ is related to the multifractality of the single particle wave function. As discussed in Ref.~\onlinecite{1992_Hiramoto_AAScaling}, the multifractal analysis of the wave function depends on whether it is done in the ``center'' of the band or near the ``edge'' close to the Van-Hove singularities, and thus we assume $d_2$ is energy dependent. 
Nonetheless, our aim is to compute $\nu$ and not the multifractal properties, therefore we find it convenient to average the IPR over suitable energy windows, which smears out the multifractal properties (in energy) and our computed value of $d_2$ is then in a sense an average over a part of the band.

\begin{table}[htp]
\begin{tabular}{| c | c | c | }
\hline  & $\nu$ & $d_2$ \\ \hline 
$\alpha = 0$  & $0.98\pm0.05$ & $0.51\pm0.03$  \\ \hline
$\lambda=\lambda_c$  & $0.97\pm0.07$ & $0.51\pm0.03$ \\ \hline
$\lambda = 0.9$ ($\epsilon >\epsilon_{ME}$)  & $0.95\pm0.08$ & $0.92\pm0.05$ \\ \hline
$\lambda = 0.9$ ($\epsilon <\epsilon_{ME}$)  & $1.05\pm0.08$ & $0.90\pm0.06$     \\ \hline 
\end{tabular}
\caption{Critical exponents of the generalized Aubry-Andr\'e model\label{tab:exponents}}
\end{table}

\begin{figure}[htp!] 
\includegraphics[angle=0,width=\linewidth]{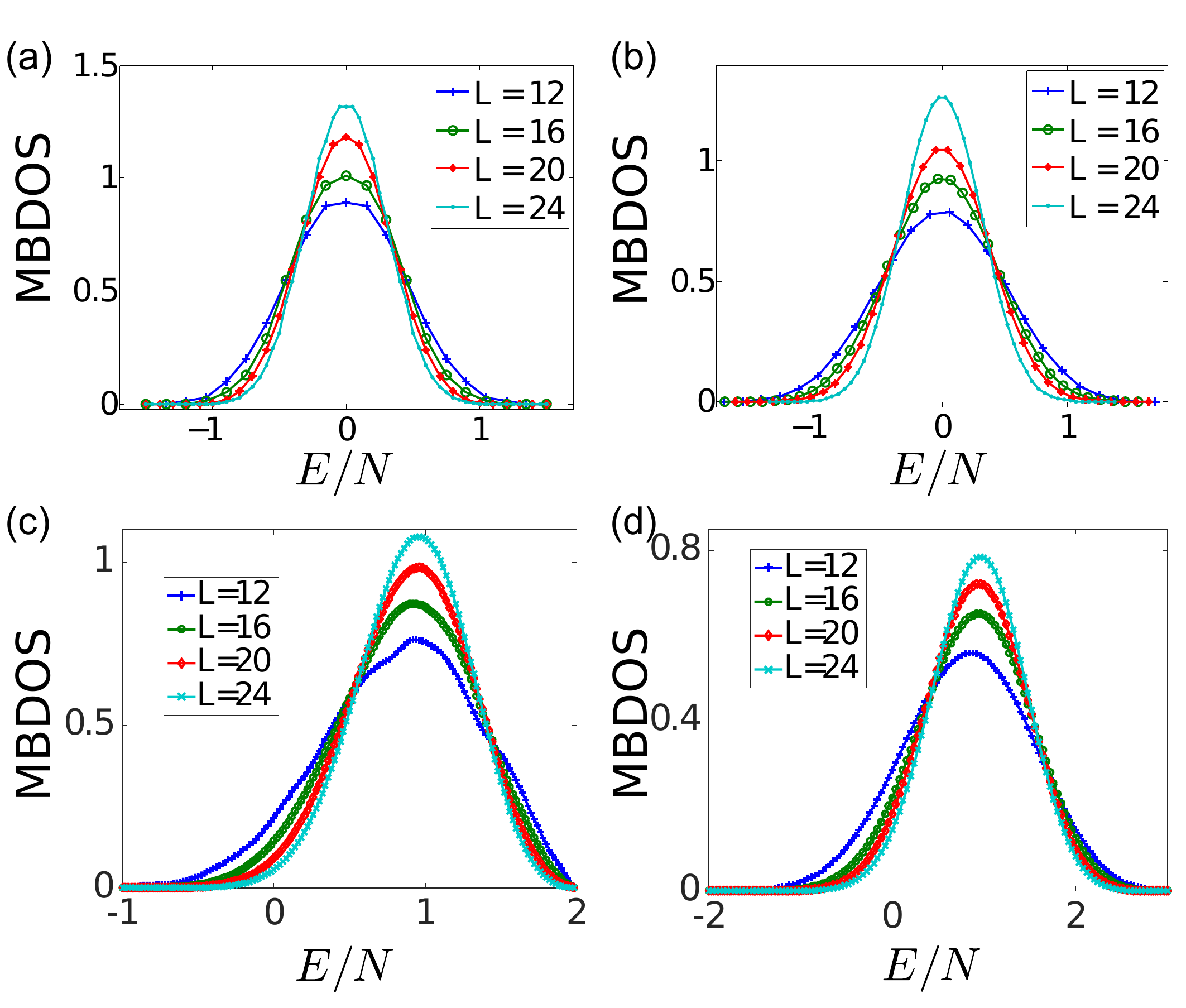}
 \caption{Many body density of states (MBDOS) of the one dimensional Anderson model at half filling. We show the MBDOS for the four different cases,  (a) without disorder or interaction, (b)  with disorder but no interaction, (c) without disorder but with interaction, and (d) with both disorder and interaction. We choose $W/t =V/t = 0$, 
 $W/t = 2, V/t = 0$, $W/t = 0, V/t = 2$, and $W/t = V/t = 2$, in (a), (b), (c), and(d), respectively.  For the disordered cases, we average over $100$ realizations. The asymmetric shape of the distribution in (c) and (d) is due to interactions (for $V\rightarrow -V$ the MBDOS is peaked at $E<0$). 
 For the interacting case, we use the kernel polynomial method~\cite{2006_Weisse_KPM_RMP}
}
\label{fig:MBDOS} 
\end{figure}

In the AA model, the localization length is known analytically to be $\xi = 1/\log (\lambda/t)$~\cite{AA}. Thus the correlation length critical exponent $\nu$ is known exactly to be $\nu = 1$. Here we first reproduce this 
analytically exact exponent purely
numerically to establish that we can compute this critical exponent to reasonably high accuracy. This accuracy will dictate how well we can 
compute the critical exponents 
for the GAA model. 
For the Aubry-Andr\'e model ($\alpha=0$), 
where there is no mobility edge, we average the IPR over all energies, with the results shown in Fig.~\ref{fig:AA_IPR} (a) and (b). 
Introducing a finite $\alpha$ (see Eq.~\eqref{eq:GAA}) gives the model a mobility edge described by~\cite{sriramgaa}
\begin{equation}
\alpha \epsilon_{ME} = 2\mathrm{sgn}(\lambda) (|t| - |\lambda|), 
\label{eq:me}
\end{equation}
 and therefore we have to carefully choose the energies so that we average over single particle orbitals that are all delocalized or all localized as to not mix in any orbitals from the other phase (Fig.~\ref{fig:GAAavescheme}).  For this we have considered two different values of $\lambda$. First we consider $\lambda = \lambda_c$ as a function of $\alpha$, due to Eq. (\ref{eq:me}) there is a mobility edge at $\epsilon=0$ and $\alpha=0$. As a result we average over all energies $\epsilon<0$ such that for $\alpha < 0$ we are capturing localized orbitals and for $\alpha>0$ delocalized orbitals (see Fig.~\ref{fig:GAAavescheme}). Thus the average IPR captures the transition as a function of $\alpha$ between a localized ($\alpha<0$) and a delocalized ($\alpha>0$) phase 
 as shown in Fig.~\ref{fig:AA_IPR} (c) and (d).  This trajectory is interesting as it keeps the Aubry-Andr\'e model as the critical point $\alpha=0$, 
 but allows us to extract the scaling from the distance to critical point as a function of $\alpha$. 
 We find the critical exponents between the AA model and the GAA model at 
 $\lambda=\lambda_c$ to be in excellent numerical agreement.
 
 We now focus on $\lambda = 0.9 t$ in the GAA model which has a mobility edge at $\epsilon_{ME} = 0.2 t/\alpha$, allowing us to study a critical point that may be distinct from the $\alpha=0$ AA limit. To this end we will consider averaging over energies $\epsilon>\epsilon_{ME}$ ($\epsilon<\epsilon_{ME}$), which for increasing $\alpha$ will give rise to a delocalized to localized  (localized to delocalized) transition at $ \alpha_c^+$ ($ \alpha_c^-$) and the results are shown in Fig.~\ref{fig:GAA_IPR}.  Here we find the critical couplings $\alpha_c^{\pm}$ from where the last eigenenergy intersects the mobility edge (Fig.~\ref{fig:GAAavescheme}(b)), for $\epsilon<\epsilon_{ME}$ this yields $\alpha_c = -0.07286$ and for $\epsilon>\epsilon_{ME}$ this yields $\alpha_c = 0.07631$. This averaging procedure and estimates of critical couplings yield excellent data collapse as seen in Fig.~\ref{fig:GAA_IPR} (b) and (d).  

From the data collapse of the average IPR we find the results listed in Table~\ref{tab:exponents}, a few comments are in order. First, interestingly we find that the critical properties of the generalized Aubry-Andr\'e model across the mobility edge matches the $\alpha=0$ limit, i.e. introducing a mobility edge by tweaking the incommensurate potential does not change the universal value of $\nu$. Second, the different energy averages do affect the value of $d_2$, but we find they do agree reasonably well for $\lambda=0.9t$ and the two different energy averages $\epsilon< \epsilon_{ME}$ and $\epsilon>\epsilon_{ME}$. Third, we find that $\nu \approx 1$ strongly violates the Chayes-Chayes-Fisher-Spencer (CCFS) bound $\nu \ge 2/d (=2)$~\cite{1986_CCFSBound_PRL} for a stable disorder driven quantum critical point. However this bound has been obtained for disordered critical points where the order parameter follows an underlying distribution. Thus, such a notion does not apply to the incommensurate models as there is no disorder (in a strict sense) and our analysis of $\nu$ places them into their own class.  This provides a major distinction between Anderson localization driven by disorder that satisfies the CCFS bound and incommensurate models that do not.
In fact, it should be emphasized that Aubry-Andr\'e type (or any other related) incommensurate localization is not an Anderson localization at all in any sense-- it is not driven by quantum interference from disorder, it is caused by energy gaps in the Schr\"odinger spectrum, and hence the concept of CCFS bound is irrelevant for AA or GAA type localization.

\section{Many-body density of states} 
\label{sec：MBDOS}
A normalized many-body density-of-states (MBDOS) is introduced as, 
\be 
{\rm MBDOS} (\varepsilon) = \frac{N}{V_H} \sum_ {q} \delta (E- E_q), 
\ee 
with $\varepsilon = E/N $  the energy per particle, $V_H$ the Hilbert space dimension, and $q$  labeling many-body eigenstates. 
As defined, the MBDOS is normalized as 
$
\int d\varepsilon \, {\rm MBDOS} (\varepsilon ) = 1. 
$ 
For a large system, MBDOS is strongly peaked at $E_{\rm middle}$, with a broadening 
 proportional to $ 1/\sqrt{N}$, which vanishes in the thermodynamic limit. 
For non-interacting fermions, the form of MBDOS is restricted by the central limit theorem to be a Gaussian, 
\be 
{\rm MBDOS} (\varepsilon)= \frac{1}{\sqrt{\pi} \sigma_\varepsilon} \exp\left[-(\varepsilon-E_{\rm middle}/N)^2/\sigma_\varepsilon ^2 \right]. 
\ee 
For interacting fermions, we find that the broadening is still greatly suppressed as we increase the system size. The  difference is that the interacting MBDOS is an asymmetric function (depending on the sign of $V$) 
instead of being a Gaussian (see Fig.~\ref{fig:MBDOS}).

\bibliographystyle{apsrev4-1}
\bibliography{references}

\end{document}